\documentclass[aps,prb,twocolumn,superscriptaddress,showpacs]{revtex4-2}

\usepackage{graphicx}
\usepackage{amsfonts}
\usepackage{amsmath}
\usepackage{amssymb}
\usepackage{bm}
\usepackage{color}

\usepackage[colorlinks, linkcolor=red,citecolor=blue,urlcolor=blue]{hyperref}
\usepackage[all]{hypcap} 
\begin{document}

\title{Adiabatic phase pumping in S/F/S hybrids with non-coplanar magnetization}

\author{A. A. Kopasov}
\affiliation{Institute for Physics of Microstructures, Russian Academy of Sciences, 603950 Nizhny Novgorod, GSP-105, Russia}
\author{Zh. Devizorova}
\affiliation{Moscow Institute of Physics and Technology (National Research University), Dolgoprudnyi, Moscow region, 141701 Russia}
\author{H. Meng}
\affiliation{School of Physics and Telecommunication Engineering, Shaanxi University of Technology, Hanzhong 723001, China}
\author{S. V. Mironov}
\affiliation{Institute for Physics of Microstructures, Russian Academy of Sciences, 603950 Nizhny Novgorod, GSP-105, Russia}
\author{A. S. Mel'nikov}
\affiliation{Institute for Physics of Microstructures, Russian Academy of Sciences, 603950 Nizhny Novgorod, GSP-105, Russia}
\affiliation{Moscow Institute of Physics and Technology (National Research University), Dolgoprudnyi, Moscow region, 141701 Russia}
\author{A. I. Buzdin}
\email{alexandre.bouzdine@u-bordeaux.fr}
\affiliation{University Bordeaux, LOMA UMR-CNRS 5798, F-33405 Talence Cedex, France}
\affiliation{World-Class Research Center ``Digital Biodesign and Personalized Healthcare'', Sechenov First Moscow State Medical University, Moscow, 19991, Russia}

\begin{abstract}
We study the distinctive features of the phase pumping effect in Josephson transport through a three-layered ferromagnet F$_1$/F/F$_2$ with non-coplanar magnetization. Using Gor'kov and Bogoliubov-de Gennes formalisms we go beyond the quasiclassical approximation and analyze the dependence of the spontaneous Josephson phase $\psi$ on the exchange field $h$ in the F layer and details of magnetization profile. The pumping of the Josephson phase can be generated by the mutual rotation of magnetizations in $F_1$ and $F_2$ layers resulting in the nontrivial phase gain at the rotation period (Berry phase). The increase in $h$ is shown to cause changes in the topology of the phase evolution: the gain of the Josephson phase at the pumping period switches from $0$ to $2\pi$. We study the scenario of these switchings originating from the interplay between several competing local minima of the free energy of the junction versus the superconducting phase difference. Our analysis provides the basis for the search of experimental setup realizing the phase pumping phenomenon.
\end{abstract}

\maketitle

\section{Introduction}

The phenomenon of quantum pumping in different mesoscopic systems attracts the interest of both experimentalists and theoreticians for several decades (see, for example,~\cite{AltshulerS1999,SwitkesS1999,MoskaletsPRB2005,ButtikerJLTP2000} and references therein). One of the first suggestions of adiabatic charge pumping mechanisms was made by D.~Thouless in his seminal paper~\cite{ThoulessPRB1983} where he considered the dynamics of quantum particles in the moving periodic potential. Being rather general, the idea of adiabatic pumping can be naturally applied not only for a charge variable but also for other physical quantities. In particular, the superconducting Josephson-type systems are known to provide an interesting possibility to realize the Thouless pumping scenario for the superconducting phase variable~\cite{Nazarov_PRL}, which is dual to the electric charge (see~\cite{AstafievN2022} and references therein). To create a driving potential for the superconducting phase we need to consider the systems, which allow the continuous tuning of the equilibrium Josephson phase between the superconducting electrodes. This possibility can be realized in so-called $\varphi_0$ junctions possessing an unconventional current phase relation $I_s(\varphi)=I_c\sin(\varphi-\varphi_0)$ and revealing, thus, a spontaneous phase difference $\varphi_0\neq \{0,~\pi\}$ in the ground state. The appearance of a nonzero spontaneous phase is possible only for Josephson systems with broken time-reversal and inversion symmetries~\cite{Buzdin}. Being integrated into the superconducting loop such $\varphi_0$ junction should produce spontaneous electric current~\cite{Ustinov, Bauer, Buzdin_2005, Feofanov, Ortlepp}. This anomalous Josephson effect arises in a variety of systems involving unconventional superconductors~\cite{Geshkenbein, Yip, Sigrist, Kashiwaya}, topological insulators~\cite{Tanaka_TI, Houzet_TI, Aubin}, Josephson junctions consisting of conventional superconductors separated by magnetic normal metal~\cite{Buzdin, Reynoso, Mironov_SOC, Bergeret_SOC}, quantum dot~\cite{Martin, Martin_2, Brunetti} or semiconductor nanowire with strong spin-orbit interaction~\cite{Nazarov_PRB, Campagnano, Kouwenhoven, Nesterov, Ying, Spanslatt, Kutlin, Kopasov}. In most cases the spontaneous phase can be tuned by varying a certain parameter, however the typical range of such tuning appears to be strongly restricted. For instance, in a large class of Josephson systems with the spin-orbit interaction the spontaneous phase can be tuned by changing the direction of the exchange (or Zeeman) field $\mathbf{h}$ with the characteristic variation range of the anomalous phase restricted by the value $\sim\alpha h L$ which is typically small (here $\alpha$ is the spin-orbit constant and $L$ is the junction length). Restrictions on the possible range of the anomalous phase can also be posed by the system design. Prominent examples include the superconductor/ferromagnet/superconductor (S/F/S) Josephson junctions with varying thickness of the F layer (see Refs.~\cite{GurlichPRB2010,SickingerPRL2012} and references therein), long Josephson junctions with current injectors acting as an effective source of the phase jumps along the junction (see, e.g., Refs.~\cite{UstinovAPL2002,GaberPRB2005,GoldobinPRB2016}) as well as curved nanowire junctions~\cite{Ying,Spanslatt,Kutlin,Kopasov}. Despite the possibility of engineering any value of the sponteneous phase in such systems, it is extremely hard to tune this phase after the system is fabricated.

\begin{figure}[t!]
	\includegraphics[width=0.35\textwidth]{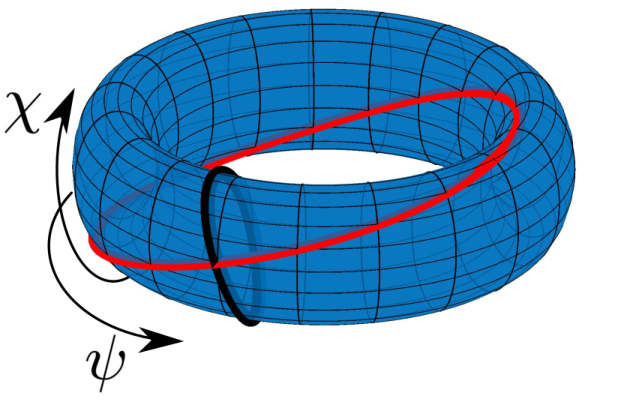}
	\caption{Schematic picture of the parameter space for S/F$_1$/F/F$_2$/S Josephson junctions with non-coplanar magnetization. Here $\psi$ is the spontaneous Josephson phase and $\chi$ is the angle between projections of the exchange fields in F$_1$ and F$_2$ to the plane perpendicular to the exchange field in F. Red (black) line shows a nontrivial (trivial) trajectory characterized by the nontrivial (trivial) gain of the superconducting phase at the rotation period.}
\label{Fig:torus_figure}
\end{figure}

The continuous tuning of the spontaneous phase can be realized in the Josephson junction with a weak link consisting of half-metal (HM)~\cite{Pickett, Coey, Keizer, Anwar} and surrounded by two conventional ferromagnets $F_1$ and $F_2$. To stress this difference between conventional $\varphi_0$ junctions and structures where the spontaneous phase can be tuned in the whole range between $0$ and $2\pi$ we will refer to the latter systems as $\psi$ junctions. The key ingredient for such tuning is the non-coplanar magnetization distribution which provides a phase bias for equal-spin Cooper pairs determined by the mutual orientations of magnetic moments in F$_1$ and F$_2$ layers. The pumping of the Josephson phase can be generated by the mutual rotation of magnetizations in side ferromagnets resulting in the nontrivial phase gain at the rotation period~\cite{Nazarov_PRL, Eschrig_1, Eschrig_2, Eschrig_3, Eschrig_4, Mironov_HM, Meng_Flux} (see the thick red line in Fig.~\ref{Fig:torus_figure}). Several important results regarding the behavior of the spontaneous phase in Josephson systems with non-coplanar magnetization were obtained in Ref.~\cite{Nazarov_PRL} within the framework of the circuit theory. In particular, it was demonstrated that for the S/F$_1$/HM/F$_2$/S structure an equilibrium superconducting phase difference is not small and continuously depends on the magnetic configuration. For S/F$_1$/F/F$_2$/S system it was shown that the interference of the contributions from equal-spin Cooper pairs to the Josephson current can cause $0$-$\pi$ transition upon the change in the magnetic configuration. It is important to note that the circuit theory analysis in Ref.~\cite{Nazarov_PRL} relies on the quasiclassical approximation (which is applicable only for small exchange field values) and is relevant only for rather long diffusive junctions, so that the length of the junction greatly exceeds the spatial scale of the spin-singlet and short-ranged spin-triplet superconducting correlations. One can naturally expect that for rather short junctions the behavior of the spontaneous phase can be qualitatively different from the one offered by a circuit theory. Indeed, in this case one can't exclude the contributions to the Josephson current from the spin-singlet and short-ranged spin-triplet Cooper pairs. Being sensitive to the magnitude of the spin polarization in the central F layer, these additional contributions can, in turn, affect the current-phase relation of the junction and the value of the anomalous phase. There are several important questions that remain open for both clean and diffusive Josephson systems with non-coplanar magnetization. These questions relate to the behavior of the ground-state superconducting phase difference upon the change in a magnetic configuration in S/F$_1$/F/F$_2$/S junctions for arbitrary ratio of the Fermi energy and the exchange field $h$ in the central F layer. Keeping in mind that the phase pumping effect is absent for $h = 0$, one can naturally expect that the increase in $h$ should cause changes in the topology of the phase evolution: the gain of the Josephson phase at the pumping period should switch from 0 to $2\pi$ (see Fig.~\ref{Fig:torus_figure}). The analysis of the above problems is known to require a theoretical approach, which goes beyond the standard quasiclassical approximation in the ferromagnet (for details see Ref.~\cite{Devizorova}).

In the present paper we fill these gaps and develop a theory of the anomalous Josephson effect in S/F$_1$/F/F$_2$/S junctions beyond the quasiclassical approximation. For such structures we demonstrate the possibility to create $\psi$ junction in which the Josephson phase can be tuned by  varying the direction of magnetization in F$_1$ or F$_2$ layer, e.g., under the effect of external magnetic field. To highlight the main qualitative features of the $\psi$ junctions we first consider a models of atomically thin SF/HM/SF Josephson junction, (see Fig.~\ref{Structure_Atomic}) where the leads consists of ferromagnetic superconductors (SF) with the built-in exchange field, the electron motion in the plane of the layers is assumed to be ballistic while the electron transfer across the structure is associated with the tunneling processes. Using the combination of the microscopical Gor'kov formalism and tight-binding model we calculate the critical current and spontaneous Josephson phase $\psi$ of the junction. Although such model does not account the interference effects coming from the finite thickness of the layers, it allows the exact analytical solution which does not rely on any sort of quasiclassical approaches or numerical modeling. The results of calculations clearly show that non-coplanarity in the magnetic configuration causes the generation of $\psi$ junction where the spontaneous phase $\psi$ equals to the angle between projections of the exchange fields in the SFs layers to the plane perpendicular to the exchange field of the central half-metallic layer. Specifically, $\sin(\psi) =  {\bf n}_h\cdot({\bf n}_1\times \bf {n}_2)/\left|\sin\theta_1\sin\theta_2\right|$ where ${\bf n}_1$, ${\bf n}_2$ and ${\bf n}_h$ are the unit vectors along the exchange field in the F$_1$, F$_2$ and half-metallic F layer, respectively, while $\theta_1$ and $\theta_2$ are the angles between the vectors ${\bf n}_1$, ${\bf n}_2$ and the spin quantization axis ${\bf n}_h$ in the half-metal.

We then consider clean three-dimensional S/F$_1$/F/F$_2$/S junctions with a finite thickness of the central layer and non-coplanar magnetization distribution. For simplicity, the effects of the finite thickness of the side ferromagnets F$_1$ and F$_2$ are neglected. Correspondingly, their role in the Josephson transport is reduced to the spin-active boundary conditions for the quasiparticle wave function. Several properties of similar hybrid structures were addressed in Refs.~\cite{Margaris,Halterman}. The current-phase relation and the behavior of the supercurrent at zero superconducting phase difference (the so-called anomalous current) as a function of hybrid structure parameters were analyzed in Ref.~\cite{Margaris} for the one-dimensional S/F/S Josephson junctions with the spin-active interfaces. Extensive analysis of various characteristics of clean three-dimensional S/F$_1$/F/F$_2$/S junctions including the current-phase relation, the anomalous current, the spatial profiles of the pair amplitude and the density of states was performed in Ref.~\cite{Halterman}. The focus of the present work is on the behavior of the spontaneous superconducting phase difference $\psi$ and the distinctive features of the phase pumping effect. The calculations of the Josephson transport are carried out within the framework of the theoretical approach used in Ref~\cite{KalenkovPRB2009}. For this purpose, we derive the Bogoliubov - de Gennes (BdG) equations for the hybrid structure and then express the supercurrent in terms of the determinant of the matching condition matrix. Such an approach is known to be equivalent to the standard one~\cite{BeenakkerPRL1991} and is suitable for the Josephson junctions of arbitrary length. It is shown that if the exchange field $h$ in F layer exceeds the Fermi energy (F is a half-metal) the spontaneous phase $\psi$ is proportional to the angle between projections of the exchange fields in F$_1$ and F$_2$ to the plane perpendicular to the exchange field in F. It is demonstrated that when decreasing the $h$ value below the Fermi energy the free energy of the junction as a function of the superconducting phase has two competing local minima resulting in jump-wise changes of the spontaneous phase $\psi$ upon magnetization rotation accompanied with the hysteresis phenomena. Further decrease in $h$ is shown to induce several changes in the topology of the phase evolution: the gain of the Josephson phase at the rotation period switches between 0 to $2\pi$ (compare black and red trajectories in Fig.~\ref{Fig:torus_figure}). It is demonstrated that the tunability of the spontaneous phase as a function of the relative orientation of the magnetic moments in three ferromagnetic layers can persist up to rather small exchange fields in the central one. Our numerical results also reveal rather complex behavior of the ground-state superconducting phase difference as a function of the structure parameters. In particular, in our simulations performed for short ballistic junctions we observe a prominent role of the size quantization effects in the behavior of the spontaneous phase, which manifest themselves through oscillatory and/or the jump-wise behavior of the anomalous phase upon the change in the exchange field or the junction length. All the results of our numerical simulations are clarified by the analytical expression for the current-phase relation, which has been derived for the case of a large mismatch between the Fermi velocities in the superconducting leads and in the central ferromagnet.

The paper has the following structure. In Sec.~\ref{Sec_atomic} using the exact solution of Gor'kov equations we analyze the current-phase relation of the atomically thin SF/HM/SF junction. Sec.~\ref{Sec_BdG} is devoted to the analysis of the behavior of the ground-state superconducting phase difference and the phase pumping phenomenon for clean S/F$_1$/F/F$_2$/S Josephson junctions with a finite thickness of the central layer. In Sec.~\ref{Sec_conc} we summarize our results.

\section{SF/HM/SF Josephson junction of atomic thickness}\label{Sec_atomic}

\begin{figure}[t!]
	\includegraphics[width=0.48\textwidth]{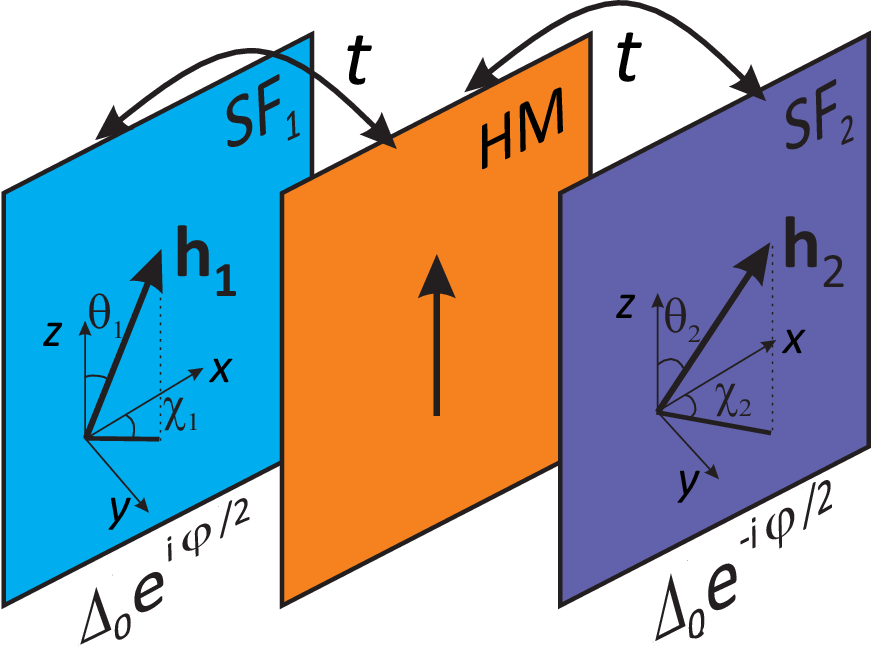}
	\caption{Sketch of the Josephson junction consisting of two superconducting ferromagnets (SFs) of atomic thickness separated by atomically thin half-metalic (HM) layer. The spin quantization axis in half-metal coincides with the $z$-axis. The layers are coupled by the transfer integrals $t$.}  \label{Structure_Atomic}
\end{figure}

In this section we analyze the current-phase relation of the Josephson junction based on the  SF$_1$/HM/SF$_2$ structure of atomic thickness. Our goal here is to find the exact solutions of the Gor'kov equations beyond the quasiclassical approximation and establish the generic conditions required for the formation of the Josephson $\psi$ junction. The system geometry is schematically shown in Fig.~\ref{Structure_Atomic}. The $y$-axis is chosen perpendicular to the layers' interfaces. The spin quantization axis in the HM layer coincides with the $z$ axis, while the exchange field ${\bf h}_j$ in the SF$_j$ layer forms the angle $\theta_j$ with the $z$ axis and the angle $\chi_j$ with the $x$-axis in the $xy$-plane: ${\bf h}_j=h(\sin \theta_j \cos \chi_j,\sin \theta_j \sin \chi_j, \cos \theta_j)$, where $j=1,2$. We assume the electron motion to be characterized by the momentum ${\bf p}$ in the plane of each layer while the quasiparticles transfer perpendicular to the layers is described by the tight-binding model. The transfer integral $t$ between SF$_{j}$  and HM layers is assumed to be much smaller than the critical temperature $T_c$. Also we restrict ourselves to the limit of coherent interlayer tunneling which conserves the in-plane electron momentum. The in-plane quasiparticles motion in the SF layers is described by the energy spectrum $\xi({\bf p})$, while the energy spectrum in the half-metal is spin dependent: $\xi_{\uparrow}=\xi({\bf p})$ and $\xi_{\downarrow}=+\infty$. The gap functions are $\Delta_1=\Delta_0 e^{ i \varphi/2}$ and $\Delta_2=\Delta_0 e^{ -i \varphi/2}$. We follow the approach of \cite{Devizorova_2} and introduce the electron annihilation operators $\phi$, $\zeta$ and $\eta$ in the SF$_1$, HM and SF$_2$ layers, respectively. Then the system Hamiltonian can be written in the following form: 
\begin{equation}\label{H_gen}
\hat H=\hat H_0+\hat H_{BCS}+\hat H_{t},
\end{equation}
where the operator 
\begin{equation}
\label{H0}
\hat H_0=\sum \limits_{{\bf p};\alpha,\beta=\{\uparrow,\downarrow\}} \left(\hat A^{(1)}_{\alpha \beta }\phi^\dag_{\alpha}\phi_{\beta}+\hat P_{\alpha \beta }\zeta^\dag_{\alpha}\zeta_{ \beta}  +\hat A^{(2)}_{\alpha \beta }\eta^\dag_{\alpha}\eta_{\beta}\right)
\end{equation}
describes the electron motion in the isolated layers ($\alpha$ and $\beta$ are the spin indexes), the contribution
\begin{multline}
\label{H_BCS}
\hat H_{BCS}=\sum \limits_{{\bf p}}\biggl(\Delta^*_1  \phi_{{-\bf p},\downarrow}\phi_{{\bf p},\uparrow}+\Delta_1  \phi^\dag_{{\bf p}, \uparrow}\phi^\dag_{-{\bf p}, \downarrow}   +\\+\Delta^*_2  \eta_{{-\bf p},\downarrow}\eta_{{\bf p},\uparrow}+\Delta_2  \eta^\dag_{{\bf p}, \uparrow}\eta^\dag_{-{\bf p}, \downarrow}\biggr),
\end{multline}
stands for the s-wave superconducting coupling inside the junction electrodes, the operator
\begin{multline}
\label{Ht}
\hat H_t=\sum \limits_{n, {\bf p}, \alpha=\{\uparrow,\downarrow\}} t (\phi^\dag_{{\bf p}, \alpha} \zeta_{{\bf p}, \alpha} + \zeta^\dag_{{\bf p}, \alpha} \phi_{{\bf p} \alpha} ) +\\+t (\zeta^\dag_{{\bf p}, \alpha} \eta_{{\bf p}, \alpha} + \eta^\dag_{{\bf p}, \alpha} \zeta_{{\bf p} \alpha} ),
\end{multline}
reflects the coherent electron tunneling between the layers, and the matrices $\hat A^{(j)}$ and $\hat P$ are defined as
\begin{equation}
\label{A}
\hat A^{(j)}=\left( \begin{array}{cc}
\xi({\bf p})-h\cos \theta_j & -h e^{-i \chi_j }\sin \theta_j\\
-he^{i \chi_j }\sin \theta_j & \xi({\bf p})+h\cos \theta_j
\end{array} \right),
\end{equation}
\begin{equation}
\label{P}
\hat P=\left( \begin{array}{cc}
\xi({\bf p}) & 0\\
0 & +\infty
\end{array} \right).
\end{equation}

The appearance of the spontaneous ground-state Josephson phase $\psi$ and its direct relation to the relative orientation of the exchange fields in SF$_1$ and SF$_2$ layers (namely, $\psi=\chi_2-\chi_1$) straightly follows from the form of the Hamiltonian (\ref{H_gen}). Indeed, below we show that it is possible to make the simultaneous phase transformation of all annihilation operators which does not change the form of the Hamiltonian (\ref{H_gen}) but makes two modifications. First, this transfromation effectively rotates the exchange field vectors in the $xy$ plane in a way that the angle between their projection to this plane becomes zero. Second, it produces an additional phase shift $\psi$ between the superconducting electrodes of the Josephson junction. The appearance of this spontaneous phase reflects the anomalous Josephson effect.

The generic phase transformation of all annihilation operators reads $\phi_{\alpha}=\tilde{\phi}_{\alpha} e^{i\kappa_{\alpha}}$, $\zeta_{\alpha}=\tilde{\zeta}_{\alpha}e^{i\mu_{\alpha}}$, $\eta_{\alpha}=\tilde{\eta}_{\alpha}e^{i\nu_{\alpha}}$, where $\kappa_{\alpha}$, $\mu_{\alpha}$ and $\nu_{\alpha}$ are certain constants. Since we assume the full spin polarization of electrons inside the half-metal the spin-down component of the corresponding annihilation operator $\zeta_\downarrow=0$ and, thus, the part $\hat H_t$ of the Hamiltonian does not change, if we take $\kappa_1=\mu_1$ and $\mu_1=\nu_1$. Moreover, if in addition we choose $\kappa_2-\kappa_1=\chi_1$ and $\nu_2-\nu_1=\chi_2$ the part $\hat H_0$ remains the same as in Eq.~(\ref{H0}), but with the modified form of the matrix $\hat A^{j}$: $\hat A^{j} \rightarrow \tilde{A}^{j}=\xi({\bf p})\sigma_0 -h \cos \theta_{j} \sigma_z -h \sin \theta_{j} \sigma_x$. The form of the transformation for the matrix $\hat A^{j}$ makes the problems equivalent to the case of the exchange fields lying in the $xz$-plane. Also the described transformation of $\hat A^{j}$ is accompanied by the corresponding modification of the order parameter values $\Delta_1$ and $\Delta_2$ in the superconducting part $H_{BCS}$ of the Hamiltonian: $\Delta_1 \rightarrow \tilde{\Delta}_1=\Delta_1 e^{-i(\kappa_1+\kappa_2)}$  and $\Delta_2 \rightarrow \tilde{\Delta}_2=\Delta_2 e^{-i(\nu_1+\nu_2)}$. Since the exchange field lying in the $xz$ plane does not affect the Josephson phase, we find: ${\rm Arg}(\tilde{\Delta}_1) -{\rm Arg}(\tilde{\Delta}_2)=\varphi +\chi_2-\chi_1$. From this expression one sees that even for $\varphi=0$ the superconducting leads effectively acquire the announced phase difference  
\begin{equation}
\psi=\chi_2-\chi_1,
\end{equation}
which corresponds to the current-phase relation
\begin{equation}
I=I_c\sin\left(\varphi+\psi\right).
\end{equation}
Note that in the case when the exchange field in the central layer is comparable or less than Fermi energy (instead of half metal we have a conventional ferromagnet) one cannot perform the above procedure. Indeed, since in this case $\zeta_\downarrow \ne 0$, one should also put $\kappa_2=\mu_2=\nu_2$ to leave the operator $\hat H_t$ unchanged. But in this case it is impossible to choose $\kappa_2-\kappa_1=\chi_1$ and $\nu_2-\nu_1=\chi_2$. Thus, the formation of the $\psi$ junction with the spontaneous ground state phase defined {\it only} by the mutual orientation of magnetic moments in ferromagnetic layers requires the material with the full spin polarization. The emergence of the spontaneous phase in S/F$_1$/F/F$_2$/S systems where the spin polarization in the central ferromagnetic layer is not full is treated in detail in Sec.~\ref{Sec_BdG}.

\begin{figure}[t!]
	\includegraphics[width=0.48\textwidth]{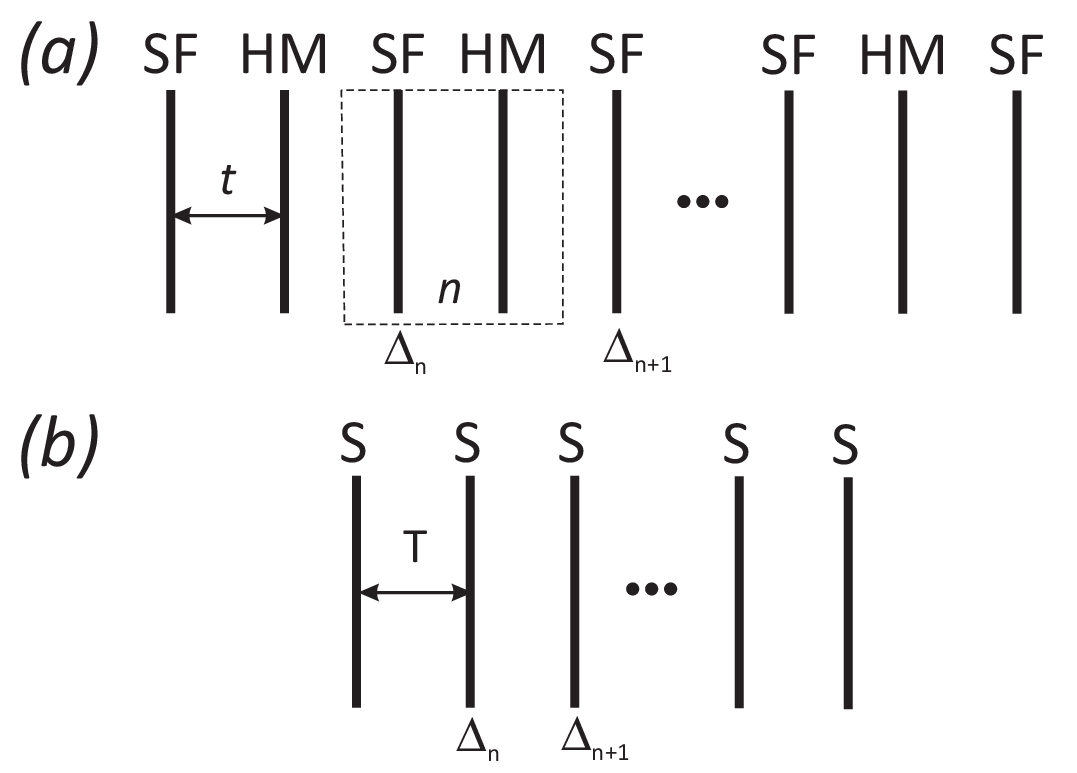}
	\caption{(a) The sketch of infinite multilayered structure consisting of alternating superconducting ferromagnets and half-metals of atomic thickness. The layers are coupled by the transfer integrals $t$, while the gap varies from one unit cell $n$ to another as $\Delta_n=\Delta_0 e^{ikn}$. This structure is equivivalent to the one shown in (b), where superconducting layers are coupled by the effective parameter ${\mathcal T}$. }  \label{Multilayers}
\end{figure}

After establishing the appearance of the spontaneous Josephson phase we now turn to the analysis of the critical current $I_c$ for the SF/HM/SF junction. The value $I_c$ depends on the angle $\theta$ between the magnetizations in the SF layers and the spin quantization axis in half-metal (without loss of generality we assume $\theta_1=\theta_2=\theta$) and is nonzero for $\theta \ne 0$, $\pi$. To show this, let us calculate this current. The above analysis shows that $I_c$ for the Josephson junction under consideration is the same as the critical current for SF/HM/SF structure with the exchange fields of the SF layers both lying in the $xz$ plane and forming the angle $\theta$ with $z$-axis. To calculate the Josephson energy of the later junction we consider the infinite multilayered structure consisting of alternating atomically thin SF (the exchange field is ${\bf h}=h\cos \theta \hat{\bf z}+h\sin \theta \hat{\bf x}$) and HM (the spin quantization axis coincides with the $z$ axis) layers and the gap varying from one unit cell to another as  $\Delta_n=\Delta_0 e^{ikn}$ (see Fig.~{\ref{Multilayers}}a). Introducing the effective coupling constant ${\mathcal T}$ between the superconducting layers (see Fig.~{\ref{Multilayers}}b) we can write down the superconducting contribution to the free energy for such SF/HM structure as:

\begin{multline}
\label{Fs}
F_s=\sum\limits_{n}{\left[\tau {\mathcal N}(0)|{{\Delta }_{n}}{{|}^{2}}+{\mathcal T} {\mathcal N}(0)(\Delta_{n}\Delta^*_{n+1}+\Delta_{n+1}\Delta^*_n) \right]}\\={{\Delta }_{0}}^{2}{\mathcal N}(0)(\tau +2{\mathcal T}\cos k)N,
\end{multline}
where $\tau=(T-T_{c0})/T_{c0}$, $T_{c0}$ is the critical temperature in the absence of the proximity effect ($t=0$) and the exchange field ($h=0$), ${\mathcal N}(0)$ is the electron density of states at the Fermi level, $N$ is the number of superconducting layers. 

From Eq.(\ref{Fs}) one sees that the Josephson energy of a single junction with the phase difference $\varphi$ between the superconducting leads reads
\begin{equation}
{{E}_{J}} = \Delta_{0}^{2}{\mathcal N}(0)\left(\tau+{\mathcal T}\cos \varphi\right) .
\end{equation}
Thus, we drive to the expression for the critical current:
\begin{equation}
\label{Ic}
I_c=-\frac{4\pi \Delta_0^2 {\mathcal N}(0) {\mathcal T}}{\Phi_0}.
\end{equation}

The coupling parameter ${\mathcal T}$ entering Eq.~(\ref{Ic}) can be expressed via the critical temperatures $T_c^0$ and $T_c^{\pi}$ for $0$-phase ($k=0$) and $\pi$-phase ($k=\pi$) of SF/HM superlattice, respectively. Indeed, in the superconducting transition we have $F_s=0$ and $\tau_c =-2{\mathcal T}\cos k$. As the result, the coupling parameter ${\mathcal T}$ reads as: $4{\mathcal T}=(T_c^{\pi}-T_c^0)/T_{c0}$.

Our next step is to find $T_c^{\pi}$ and $T_c^0$ using the microscopical Gor'kov formalism. To that end, we will write the system of matrix Gor'kov equations, find the anomalous Green function solving this system and calculate $T_c$ using the self-consistency equation. 

The Hamiltonian of the SF/HM superlattice has the form
\begin{equation}
\label{H_mult}
\hat H=\hat H_0+\hat H_{BCS}+\hat H_{t},
\end{equation}
where the three operators describe the electron motion in the plane of the layers, superconducting pairing in the superconducting layers and tunneling between the layers, respectively:
\begin{equation}
\label{H0_mult}
\hat H_0=\sum \limits_{n;{\bf p};\beta, \gamma=\{\uparrow,\downarrow\}} \left(\hat C_{\beta \gamma}\phi^\dag_{n, {\bf p}, \beta}\phi_{n, {\bf p}, \gamma}+\hat P_{\beta \gamma}\zeta^\dag_{n, {\bf p}, \beta}\zeta_{n, {\bf p}, \gamma}\right),
\end{equation}
\begin{equation}
\hat H_{BCS}=\sum \limits_{{\bf p}}\left(\Delta^*  \phi_{n, {-\bf p},\downarrow}\phi_{n,{\bf p},\uparrow}+\Delta  \phi^\dag_{n, {\bf p}, \uparrow}\phi^\dag_{n, -{\bf p}, \downarrow}\right),
\end{equation}
\begin{multline}
\label{Ht_mult}
\hat H_t=t\sum \limits_{n, {\bf p}, \beta=\{1,2\}}( \zeta^\dag_{n, {\bf p}, \beta}\psi_{n-1, {\bf p}, \beta}+\zeta^\dag_{n-1, {\bf p}, \beta}\phi_{n, {\bf p}, \beta})+\\+( \phi^\dag_{n, {\bf p}, \beta}\zeta_{n, {\bf p}, \beta}+\zeta^\dag_{n, {\bf p}, \beta}\phi_{n, {\bf p}, \beta}).
\end{multline}
Here $\phi$ and $\zeta$ are the electron annihilation operators in the SF and HM layers, respectively, $n$ is the number of a unit cell and the matrix $\hat C$ is defined as
\begin{equation}
\hat C=\left( \begin{array}{cc}
\xi({\bf p})-h\cos \theta & -h\sin \theta\\
-h\sin \theta & \xi({\bf p})+h\cos \theta
\end{array} \right).
\end{equation}

Since the multilayered system is periodic in space, we can introduce the Fourier components of the annihilation operators
\begin{equation}
\phi_{n, {\bf p}, \beta}=\int^{\pi}_{-\pi}\frac{dq}{2\pi}e^{iqn}\phi_{q, {\bf p}, \beta}, ~~ \zeta_{n, {\bf p}, \beta}=\int^{\pi}_{-\pi}\frac{dq}{2\pi}e^{iqn}\zeta_{q, {\bf p}, \beta},
\end{equation}
and write down the Hamiltonian (\ref{H_mult})-(\ref{Ht_mult}) in the Fourier representation:
\begin{equation}
\label{H0_q}
\hat H_0=\sum \limits_{q;{\bf p};\beta, \gamma=\{\uparrow,\downarrow\}} \left(\hat C_{\beta \gamma}\phi^\dag_{q, {\bf p}, \beta}\phi_{q, {\bf p}, \gamma}+\hat P_{\beta \gamma}\zeta^\dag_{q, {\bf p}, \beta}\zeta_{q, {\bf p}, \gamma}\right),
\end{equation}
\begin{equation}
\label{H_BCS_q}
\hat H_{BCS}=\Delta_0\sum \limits_{q,{\bf p}}\left( \phi_{k-q, {-\bf p},\downarrow}\phi_{q,{\bf p},\uparrow}+\Delta  \phi^\dag_{q, {\bf p}, \uparrow}\phi^\dag_{k-q, -{\bf p}, \downarrow}\right),
\end{equation}
\begin{equation}
\label{Ht_q}
\hat H_t=\sum \limits_{q, {\bf p}, \beta=\{\uparrow,\downarrow\}} (T(q)\phi^\dag_{q, {\bf p}, \beta}\zeta_{q, {\bf p}, \beta}+ T^*(q)\zeta^\dag_{q, {\bf p}, \beta}\phi_{q, {\bf p}, \beta}),
\end{equation}
where $T(q)=2te^{iq/2} \cos (q/2)$.

Next we introduce the set of imaginary-time Green functions:
\begin{gather*}
\label{F_Fourier}
F^{\dagger}_{\alpha \beta}({\bf p},q; \tau_1, \tau_2)=\langle T_{\tau} \phi^\dag_{-q, -{\bf p}, \alpha}(\tau_1) \phi^\dag_{k+q, {\bf p}, \beta}(\tau_2)\rangle, \\
G_{\alpha \beta}({\bf p}, q; \tau_1, \tau_2)=-\langle T_{\tau} \phi_{q, {\bf p}, \alpha}(\tau_1) \phi^\dag_{q, {\bf p}, \beta}(\tau_2)\rangle, \\
F^{\zeta \dagger}_{\alpha \beta}({\bf p}, q; \tau_1, \tau_2)=\langle T_{\tau} \psi^\dag_{-q, -{\bf p}, \alpha}(\tau_1) \phi^\dag_{k+q, {\bf p}, \beta}(\tau_2)\rangle, \\
\label{E_Fourier}
E_{\alpha \beta}( {\bf p}, q; \tau_1, \tau_2)=-\langle T_{\tau} \zeta_{q, {\bf p}, \alpha}(\tau_1) \zeta^\dag_{q, {\bf p}, \beta}(\tau_2)\rangle.
\end{gather*}
The resulting system of the matrix Gor'kov equations in the frequency representation takes the following form:
\begin{gather}
\label{SGE}
(i\omega_n-\hat C)\hat G+\Delta_0 I \hat F^{\dagger}+T(k+q)\hat E=\hat 1, \\
(i\omega_n+\hat C)\hat F^{\dagger} -\Delta_0 I \hat G-T(q)\hat F^{\zeta \dagger}=0, \\
(i\omega_n-\hat P)\hat E+T^*(k+q)\hat G=0, \\
(i\omega_n+\hat P)\hat F^{\zeta \dagger} -T^*(q)\hat F^{\dagger}=0,
\end{gather}
where $\omega_n=\pi T(2n+1)$ are the fermion Matsubara frequencies and $\hat I=i \sigma_y$. Solving the system of equations (\ref{SGE}) in the lowest order over the gap potential we find the anomalous Green function:
\begin{multline}
\label{AGF_Tc}
\hat F^{\dagger}=\Delta_0\left[(i\omega_n +\hat C)-|T(q)|^2(i\omega_n +\hat P)^{-1}\right]^{-1}\hat I \\ \times \left[(i\omega_n -\hat C)-|T(k+q)|^2(i\omega_n -\hat P)^{-1}\right]^{-1}.
\end{multline}

The critical temperature $T_c$ can be expressed via $\hat F^{\dagger}_{\uparrow\downarrow}$ using the self-consistency equation:
\begin{equation}
\label{SCE_Tc_general}
{\rm ln}\left(\frac{T_{c0}}{T_c} \right)=T_c \sum_{n=-\infty}^{\infty} \left[\int^{+\infty}_{-\infty} d \xi \int^{\pi}_{-\pi}\frac{dq}{2\pi} \frac{\hat F^{\dagger}_{\uparrow\downarrow}}{\Delta_0} +\frac{\pi}{\omega_n} \right],
\end{equation}
where $T_{c0}$ is the critical temperature in the absence of proximity effect ($t=0$) and the exchange field ($h=0$).

To simplify the further calculations, we assume $t$ to be small and perform the power expansion of (\ref{AGF_Tc}) over this parameter. Next we substitute this expansion into Eq.(\ref{SCE_Tc_general}) and obtain rather cumbersome equation for $T_c$ (for details see Appendix \ref{AppendixA}). Solving this equation, we can represent the critical temperature as $T_c =\alpha -\beta \cos k$ [see Eqs.(\ref{App_Tc})-(\ref{App_b})], where
\begin{equation}
\beta=\sum_{n>0} \frac{\pi \tilde{T}_{c0}^2t^4h^2(h^4+35h^2 \tilde\omega_n^2 +70 \tilde\omega_n^4)\sin^2\theta}{W\tilde\omega_n(\tilde\omega_n^2+h^2)^3 (4\tilde\omega_n^2+h^2)^2}.
\end{equation}
Here $\tilde\omega_n=\pi {\tilde T}_{c0} (2n+1)$, $\tilde{T}_{c0}$ is the critical temperature in the absence of the proximity effect ($t=0$) and 
\begin{equation}
W=1-\sum_{n>0}\frac{4\pi {\tilde T}_{c0} h^2 \tilde\omega_n}{(\tilde\omega_n^2+h^2)^2}>0.
\end{equation}
As a result, we can easily calculate the desired difference $(T_c^{\pi}-T_c^0)=2\beta$ and the critical current $I_c=2\pi \Delta_0^2N(0) \beta/(\Phi_0 T_{c0})$. As expected, we obtain $I_c \propto \sin^2\theta$.

\begin{figure}[t!]
\includegraphics[width=0.45\textwidth]{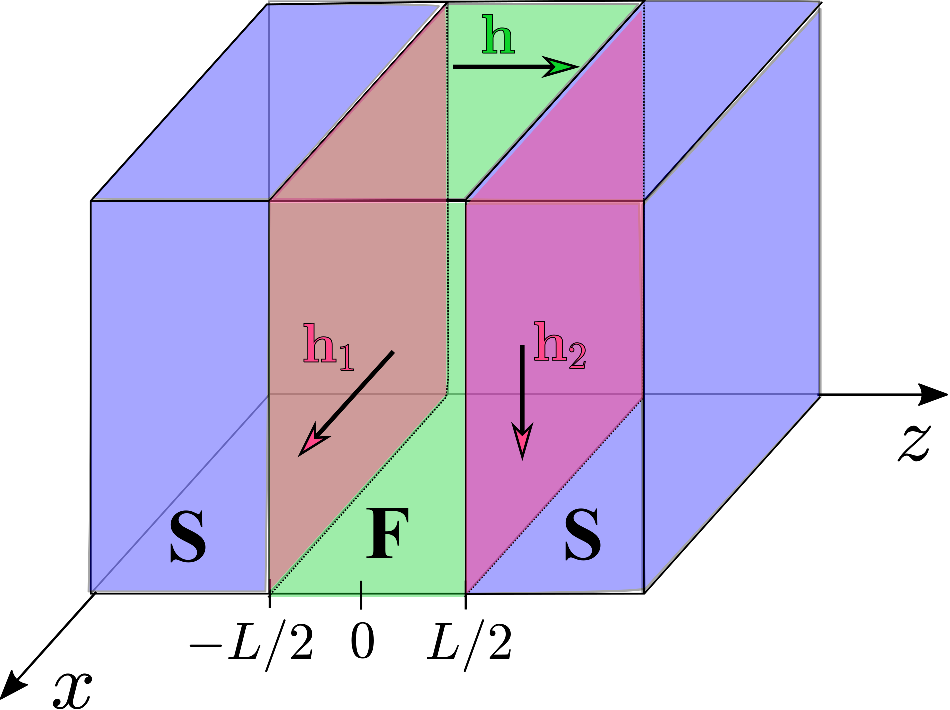}
\caption{Schematic picture of the considered S/F$_1$/F/F$_2$/S Josephson junctions with the spin-active interfaces. The exchange field ${\bf h}$ in the central ferromagnet F is perpendicular to the layers while the exchange field vectors ${\bf h}_1$ and ${\bf h}_2$ at the spin-active interfaces are parallel to the layers [see Eqs.~(\ref{exchange_field_profile})].} 
\label{Fig:model_system_finite_thickness}
\end{figure}

\section{S/F$_1$/F/F$_2$/S junctions with a finite thickness of the F layer}\label{Sec_BdG}

In this section we analyze the behavior of the spontaneous ground-state phase difference in clean S/F$_1$/F/F$_2$/S Josephson junctions with a finite thickness of the central layer $L$. Using the solutions of the Bogoliubov-de Gennes equations, we describe the crossover from the regime of half-metallic F layer to the case when the exchange field in the F layer is small compared to the Fermi energy. To simplify the calculations we treat the F$_1$ and F$_2$ layers as spin-active interfaces between the superconducting leads and the central ferromagnet. Also we assume that these interfaces are characterized by the finite potential barrier which partially damps the electron transfer between the layers.

\subsection{Geometry and basic equations}

The geometry of the system is schematically shown in Fig.~\ref{Fig:model_system_finite_thickness}. The superconducting gap is taken in the form $\Delta = \Delta_0e^{i\varphi/2}$ in the left electrode and $\Delta = \Delta_0e^{-i\varphi/2}$ in the right one. We introduce the $z$ axis along the exchange field $\mathbf{h}$ in the central ferromagnet F which is assumed to be perpendicular to the plane of the layers. The Hamiltonian of the hybrid structure reads \cite{Buz, PGdeGennes}
\begin{equation}\label{ham}
\hat{H}=\hat{H}_0+\hat{H}_{ex}+\hat{H}_{BCS} \ ,
\end{equation}
where 
\begin{equation}\label{H0}
\hat{H}_0= \int  \psi_{\alpha }^{\dagger }({\bf r})\left[-\nabla\frac{1}{2m(\mathbf{r})}\nabla-\mu(\mathbf{r}) + U(\mathbf{r})\right]\psi_{\alpha }({\bf r})d^{3}{\bf r} \ ,
\end{equation}
and the potential $U(\mathbf{r})$ is nonzero only at the interfaces between the F layer and superconducting leads:
\begin{equation}
 U(\mathbf{r}) = U_1\delta(z + L/2) + U_2\delta(z - L/2) \ .
\end{equation}
The spin-dependent part of the single-particle Hamiltonian has the following form
\begin{equation}\label{Hex}
\hat{H}_{ex}=\int  \psi_{\alpha }^{\dagger }({\bf r})\left[{\bf h}({\bf r})\cdot\hat{\boldsymbol \sigma}\right]_{\alpha\beta}\psi_{\beta}({\bf r})d^{3}{\bf r}
\end{equation}
with
\begin{equation}\label{exchange_field_profile}
\begin{array}{c}{
 \mathbf{h}(\mathbf{r}) = h_1(\mathbf{n}_1\hat{\boldsymbol{\sigma}})\delta(z + L/2) + h_2(\mathbf{n}_2\hat{\boldsymbol{\sigma}})\delta(z-L/2)}\\ {
 + h\hat{\sigma}_z\left[\Theta(z + L/2) - \Theta(z-L/2)\right]} 
 \end{array}
\end{equation}
and the unit vector
\begin{equation}
\mathbf{n}_{j} = (\sin\theta_j\cos\chi_j,\sin\theta_j\sin\chi_j,\cos\theta_j)
\end{equation}
directed along the exchange field vector in the $j$-th interface. The term
\begin{equation}\label{HBCS}
\hat{H}_{BCS}=\int  \left[ \psi_{\alpha }^{\dagger }({\bf r})\Delta ({\bf r})(i\hat{\sigma}_{y})_{\alpha \beta }
\psi_{\beta }^{\dagger }({\bf r})+h.c.\right]d^{3}{\bf r}
\end{equation}
stands for the superconducting paring inside the superconducting layers. In the above expressions $\psi_{\alpha}^{\dagger}$ ($\psi_{\alpha}$) is the fermionic creation (annihilation) operator, $\alpha,\beta = \uparrow,\downarrow$ denote spin degrees of freedom (summation over repeated spin indices is implied), $\hat{\sigma}_i$ ($i = x,y,z$) are the Pauli matrices acting in the spin space, $m(\mathbf{r})$ is the effective mass profile, and $\mu(\mathbf{r})$ denotes the difference between the chemical potential and the bottom of the electron energy band. To derive the Bogoliubov-de Gennes equations, we perform the standard Bogoliubov
transformation $\psi_{\alpha }({\bf r})=\sum_{n}[u_{n\alpha }({\bf r})
\gamma_{n} + v_{n\alpha }^{\ast }({\bf r})\gamma_{n}^{\dag }]$, where $\gamma_{n}^{\dag}$ ($\gamma_{n}$) is the quasiparticle creation (annihilation) operator, $u_{n\alpha }({\bf r})$ and $v_{n\alpha }({\bf r})$ are the electron and hole components of the quasiparticle wave function. The resulting equations take the form  
\begin{subequations}\label{BdG_equations}
 \begin{align}
 \check{H}_{\rm BdG}(\mathbf{r})\Psi(\mathbf{r}) = E \Psi(\mathbf{r}) \ ,\\
 \check{H}_{\rm BdG}(\mathbf{r}) = \check{\tau}_z\left[-\nabla\frac{1}{2m(\mathbf{r})}\nabla - \mu(\mathbf{r}) + U(\mathbf{r})\right] \\
 \nonumber
 + \mathbf{h}(\mathbf{r})\hat{\boldsymbol{\sigma}} + \check{\tau}_x{\rm Re}\Delta(\mathbf{r}) -\check{\tau}_y{\rm Im}\Delta(\mathbf{r}) \ ,
 \end{align}
\end{subequations}
where $\Psi(\mathbf{r}) = [u_{\uparrow}(\mathbf{r}),u_{\downarrow}(\mathbf{r}),v_{\downarrow}(\mathbf{r}),-v_{\uparrow}(\mathbf{r})]^{\rm T}$ and $\check{\tau}_i$ ($i = x,y,z$) are the Pauli matrices acting in the electron-hole space. 

For the calculations of the Josephson transport we follow the approach used in Ref.~\cite{KalenkovPRB2009}. This approach is based on the fact that the normal Matsubara Green's function of the system, which defines the supercurrent though the junction, can be expressed in terms of the determinant of the matching condition matrix for Eqs.~(\ref{BdG_equations}) at $E = i\omega_n$. Note that for the considered S/F/S junctions with spin-active interfaces the general expression for the Josephson current represents a trivial generalization of the one in Ref.~\cite{KalenkovPRB2009}. Before going into details, let us briefly outline the key points of the procedure. As a first step, we derive the solutions of Eqs.~(\ref{BdG_equations}) at the Matsubara frequencies in the whole system. Matching the resulting solutions at the interfaces, we derive the scattering matrices and the solvability condition matrix. Finally, the determinant of the solvability condition matrix is used to compute the Josephson current.

Below we list the appropriate solutions of Eqs.~(\ref{BdG_equations}) at $E = i\omega_n$ derived under the assumption of the in-plane translational symmetry for the stepwise $m(z)$ and $\mu(z)$ profiles. In what follows we assume the identical electron band structure in both superconducting leads. 
In the left superconductor ($z<-L/2$) the quasiparticle wave function reads
\begin{eqnarray}\label{left_sc_solution}
\Psi_I = e^{i\mathbf{p}_{||}\mathbf{r}}e^{\varkappa(z + L/2)}\biggl[e^{ip_{s}(z + L/2)}\begin{pmatrix}A_+\\e^{- i\varphi/2}a_+A_+\end{pmatrix} \\
\nonumber
+ e^{-ip_{s}(z + L/2)}\begin{pmatrix}A_-\\e^{-i\varphi/2}a_-A_-\end{pmatrix}\biggl] \ .
\end{eqnarray}
Here $\mathbf{p}_{||}$ is the conserved in-plane momentum, 
\begin{equation}
 a_{\pm}(\omega_n) = \frac{i\omega_n\pm i\sqrt{\omega_n^2 + \Delta_0^2}}{\Delta_0} \ ,
\end{equation}
$\varkappa = m_s\sqrt{\omega_n^2 + \Delta_0^2}/p_{s}$, $p_{s} = \sqrt{2m_s\mu_s - \mathbf{p}_{||}^2}$, $m_s$ is the effective mass for the electrons in superconducting leads, and $\mu_s$ is the difference between the chemical potential and the bottom of the electron energy band in the S layers. In the central ferromagnet ($-L/2<z<L/2$) the solution of BdG equations can be written as follows:
\begin{eqnarray}\label{ferromagnet_solution}
  \Psi_{II} = e^{i\mathbf{p}_{||}\mathbf{r}}\begin{pmatrix}\hat{Q}(z)B_+ + \hat{Q}(-z)B_-\\ \hat{\bar{Q}}(z)\bar{B}_++ \hat{\bar{Q}}(-z)\bar{B}_-\end{pmatrix} \ ,
\end{eqnarray}
where  
\begin{subequations}\label{ferromagnet_parameters}
\begin{align}
\hat{Q}(z) = {\rm diag}(e^{ ip_{\uparrow}z},e^{ i p_{\downarrow}z}) \ ,\\
\hat{\bar{Q}}(z) = {\rm diag}(e^{ i\bar{p}_{\uparrow}z},e^{ i \bar{p}_{\downarrow}z}) \ ,\\
\label{wave_numbers_ferromagnet}
p_{\uparrow,\downarrow} = \sqrt{2m_f\left(i\omega_n \mp h + \mu_f\right)  - \mathbf{p}_{||}^2} \ ,
\end{align}
\end{subequations}
$\bar{p}_{\uparrow,\downarrow}(i\omega_n,h) = p_{\uparrow,\downarrow}(-i\omega_n,-h)$, $m_f$ denotes the effective mass for the electrons in the central ferromagnet, and $\mu_f$ is the difference between the chemical potential and the bottom of the electron energy band in the central layer at $h = 0$. For a finite $h$ the bottom of the lower and higher spin-split subband in the F layer is located at $-(\mu_f + h)$ and $-(\mu_f - h)$, respectively. Finally, in the right superconductor ($z>L/2$) the quasiparticle wave function has the form:
\begin{eqnarray}\label{right_sc_solution}
\Psi_{III} = e^{i\mathbf{p}_{||}\mathbf{r}}e^{-\varkappa(z - L/2)}\biggl\{e^{ip_{s}(z - L/2)}\begin{pmatrix}C_+\\e^{i\varphi/2}a_-C_+\end{pmatrix} \\
\nonumber
+ e^{-ip_{s}(z - L/2)}\begin{pmatrix}C_-\\e^{i\varphi/2}a_+C_-\end{pmatrix}\biggl\} \ .
\end{eqnarray}
In the above solutions $A_{\pm}$, $B_{\pm}$, $\bar{B}_{\pm}$, and $C_{\pm}$ are 2$\times$1 column vectors in the spin space, which should be determined from the matching conditions for the quasiparticle wave function at the S/F interfaces ($z = \pm L/2$). It is straightforward to show that the corresponding matching conditions for the considered model~(\ref{BdG_equations}) 
\begin{equation}
   \begin{pmatrix}B_+\\ \bar{B}_+\end{pmatrix} = \check{\mathcal{K}}_1\begin{pmatrix}B_-\\ \bar{B}_-\end{pmatrix} \ , \ \ \begin{pmatrix}B_-\\ \bar{B}_-\end{pmatrix} = \check{\mathcal{K}}_2\begin{pmatrix}B_+\\ \bar{B}_+\end{pmatrix} \ ,
\end{equation}
yield the following expressions for the scattering matrices, which couple the electron and hole waves in the central ferromagnet:
\begin{equation}\label{scattering_matrices}
 \check{\mathcal{K}}_j = -\check{Q}(L/2)\left(\check{W}_j + \check{K}\right)^{-1}\left(\check{W}_j - \check{K}\right)\check{Q}(L/2) \ .
\end{equation}
Here 
\begin{subequations}
\begin{align}
 \check{W}_j = g\check{\tau}_z + iZ_{0,j} + iZ_j\check{\tau}_z\mathbf{n}_j\hat{\boldsymbol{\sigma}} \\
 \nonumber
 -f\left[\check{\tau}_x\sin(\varphi_j) + \check{\tau}_y\cos(\varphi_j)\right] \ ,\\
 g = \frac{\omega_n}{\sqrt{\omega_n^2 + \Delta_0^2}} \ ,\\
 f = \frac{\Delta_0}{\sqrt{\omega_n^2 + \Delta_0^2}} \ ,
 \end{align}
\end{subequations}
$\check{Q}$ and $\check{K}$ are diagonal matrices with the following structure in the electron-hole space $\check{X} = {\rm diag}(\hat{X},\hat{\bar{X}})$, $\hat{K} = (m_s/m_f) {\rm diag}(p_{\uparrow}/p_s, p_{\downarrow}/p_s)$, $Z_{0,j} = 2m_sU_j/p_s$, and $Z_j = 2m_sh_j/p_s$. The current-phase relation $I(\varphi)$ can be obtained from the determinant of the matching condition matrix
\begin{subequations}\label{supercurrent_matching_matrix}
\begin{align}\label{current_phase_BdG_definition}
 I_s(\varphi) = -2e\mathcal{A}T\sum_{\omega_n>0}\int\frac{d^2\mathbf{p}_{||}}{(2\pi)^2} \frac{\partial}{\partial\varphi}\ln\left|\mathcal{P}\right| \ ,\\
 \label{determinant_solvability_condition}
  \mathcal{P}(i\omega_n, \mathbf{p}_{||},\varphi) = {\rm det}\left|1 - \check{\mathcal{K}}_1\check{\mathcal{K}}_2\right| \ ,
 \end{align}
\end{subequations}
where $\mathcal{A}$ is the cross-sectional area of the junction. We calculate the spontaneous superconducting phase difference by minimizing the free energy of the junction, which up to a phase-independent constant is given by
\begin{equation}\label{BdG_junction_energy}
 F(\varphi) = -\mathcal{A}T\sum_{\omega_n>0}\int\frac{d^2\mathbf{p}_{||}}{(2\pi)^2}\ln\left|\mathcal{P}\right| \ .
\end{equation}
The resulting Eqs~(\ref{scattering_matrices}), (\ref{supercurrent_matching_matrix}), and (\ref{BdG_junction_energy}) form the basis for our analytical analysis and numerical simulations of the Josephson transport.

\subsection{Analytical results for the current-phase relation. Qualitative consideration}

Here we provide some analytical results, which clarify the behavior of the current-phase relation and the spontaneous superconducting phase difference. Note that the major challenge for obtaining a closed-form expression for the current-phase relation stems from the fact that the spin-active interfaces couple the quasiparticle states in the central ferromagnet from both spin-split subbands. Some analytical progress can be made when typical Fermi velocities in the central F layer are much smaller than the ones in the normal-metal state of the superconducting leads. In this case, one can consider the velocity ratio matrix $\check{K}$ in Eq.~(\ref{scattering_matrices}) as a perturbation and expand the determinant of the solvability condition matrix $\mathcal{P}$ up to the second-order terms. Choosing, for simplicity, equal barrier strength parameters $Z_{0,1} = Z_{0,2} = Z_0$, $Z_1 = Z_2 = Z$, we arrive at a sinusoidal current-phase relation of the following form:
\begin{widetext}
\begin{subequations}\label{current_phase_BdG_analytical}
\begin{align}\label{current_phase_BdG}
 I_s(\varphi) = (I_{\uparrow\downarrow}^+ + I_{\uparrow\downarrow}^-)\sin(\varphi) + I_{\uparrow\uparrow}\sin(\varphi - \chi) + I_{\downarrow\downarrow}\sin(\varphi + \chi) \ ,\\
 I_{\uparrow\downarrow}^+ = 4e\mathcal{A}T{\rm Re}\sum_{\omega_n>0}\int\frac{d^2\mathbf{p}_{||}}{(2\pi)^2}\frac{f^2[w^2-4g^2Z^2\cos(\theta_1)\cos(\theta_2)]}{\gamma(w^2 + 4g^2Z^2)^2}\left[\frac{p_{\uparrow}\bar{p}_{\uparrow}}{\sin(p_{\uparrow}L)\sin(\bar{p}_{\uparrow}L)} + \frac{p_{\downarrow}\bar{p}_{\downarrow}}{\sin(p_{\downarrow}L)\sin(\bar{p}_{\downarrow}L)}\right] \ ,\\
 I_{\uparrow\downarrow}^- = 4e\mathcal{A}T{\rm Re}\sum_{\omega_n>0}\int\frac{d^2\mathbf{p}_{||}}{(2\pi)^2}\frac{-4iwgZf^2\cos(\theta_+)\cos(\theta_-)}{\gamma(w^2 + 4g^2Z^2)^2}\left[\frac{p_{\uparrow}\bar{p}_{\uparrow}}{\sin(p_{\uparrow}L)\sin(\bar{p}_{\uparrow}L)} - \frac{p_{\downarrow}\bar{p}_{\downarrow}}{\sin(p_{\downarrow}L)\sin(\bar{p}_{\downarrow}L)}\right] \ ,\\
 I_{\uparrow\uparrow} = -4e\mathcal{A}T{\rm Re}\sum_{\omega_n>0}\int\frac{d^2\mathbf{p}_{||}}{(2\pi)^2} \frac{4g^2f^2Z^2\sin(\theta_1)\sin(\theta_2)}{\gamma(w^2 + 4g^2Z^2)^2}\frac{p_{\uparrow}\bar{p}_{\downarrow}}{\sin(p_{\uparrow}L)\sin(\bar{p}_{\downarrow}L)} \ ,\\
 I_{\downarrow\downarrow} = -4e\mathcal{A}T{\rm Re}\sum_{\omega_n>0}\int\frac{d^2\mathbf{p}_{||}}{(2\pi)^2}\frac{4g^2f^2Z^2\sin(\theta_1)\sin(\theta_2)}{\gamma(w^2 + 4g^2Z^2)^2}\frac{p_{\downarrow}\bar{p}_{\uparrow}}{\sin(p_{\downarrow}L)\sin(\bar{p}_{\uparrow}L)} \ .
 \end{align}
\end{subequations}
\end{widetext}
Here $w = 1 + Z_0^2 - Z^2$, $\theta_{\pm} = (\theta_1\pm\theta_2)/2$, $\chi = \chi_1 - \chi_2$, and $\gamma = m_f^2p_s^2/m_s^2$. The first two terms proportional to $\sin(\varphi)$ in the right-hand side of Eq.~(\ref{current_phase_BdG}) describe the contribution to the Josephson transport from the spin-singlet and the spin-triplet Cooper pairs with zero spin projection on the direction of the exchange field $\mathbf{h}/|\mathbf{h}|$. The remaining terms $\propto I_{\uparrow\uparrow}$ and $I_{\downarrow\downarrow}$ contain the contributions to the supercurrent from the parallel spin-triplet pairs $|\uparrow\uparrow\rangle$ and $|\downarrow\downarrow\rangle$, respectively.

 \begin{figure*}[htpb]
\centering
\includegraphics[scale = 0.58]{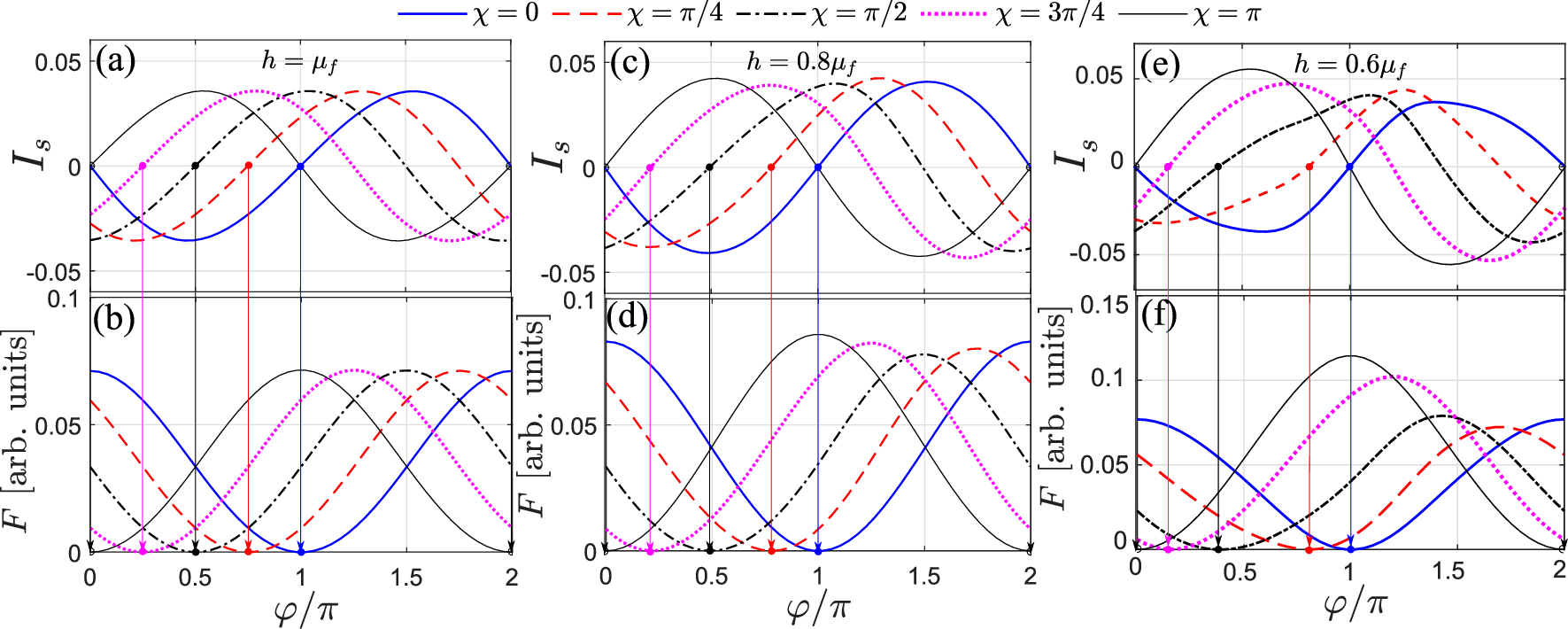}
\caption{Typical current-phase relation $I_s(\varphi)$ and $F(\varphi)$ plots for $\chi = 0$, $\pi/4$, $\pi/2$, $3\pi/4$, and $\pi$. Panels (a), (c), and (e) ((b), (d), and (f)) correspond to $h/\mu_f = 1$, $0.8$, and $0.6$, respectively. The supercurrent is given in the units $e\Delta_0\mathcal{N}$, where $\mathcal{N} = \mathcal{A}k_F^2/\pi$, $\mathcal{A}$ is the cross-sectional area of the junction, and $k_F$ is the Fermi momentum in the normal-metal state of the superconductors. Circles denote the superconducting phase difference at which the free energy of the junction reaches its minimal value (the anomalous phase $\psi$). We choose $L = 0.02\xi$ to produce the plots, where $\xi = \hbar k_F/m_s\Delta_0$ is the superconducting coherence length.
}
\label{Fig:results_BdG}
\end{figure*}

Equations~(\ref{current_phase_BdG_analytical}) provide several valuable insights into the physics of the anomalous Josephson effect in S/F$_1$/F/F$_2$/S hybrid structures. First, one can see that the above expressions capture the crossover from the S/F$_1$/F/F$_2$/S to the S/F$_1$/HM/F$_2$/S - type junctions upon the increase in the exchange field in the central ferromagnet. Indeed, for $h > \mu_f$ the contributions to the Josephson current from the higher spin-split subband in the central ferromagnet $\propto 1/\sin(p_{\uparrow}L)$, $1/\sin(\bar{p}_{\downarrow}L)$ exhibit an exponential decay and one is left with a single contribution from the lower spin-split subband
\begin{equation}
 I_{{\rm S/F/HM/F/S}}(\varphi) = I_{\downarrow\downarrow}\sin(\varphi + \chi) \ .
\end{equation}
Correspondingly, in the half-metallic regime the anomalous phase $\psi$ is a linear function of the misorientation angle $\chi = \chi_1 - \chi_2$. Second, in the limit $h\to 0$ we get $I_{\uparrow\uparrow} = I_{\downarrow\downarrow}$, which implies that the total contibution of the parallel spin-triplet Cooper pairs to the Josephson transport $\propto \cos(\chi)\sin(\varphi)$. Within this limit and for rather strong spin-active barriers one can expect that the variations of $\chi$ should cause $0$-$\pi$ transitions due to contributions of the parallel spin-triplet Cooper pairs. Note that this result is in qualitative agreement with the results of the circuit theory~\cite{Nazarov_PRL}. Third, the functional form of the current-phase relation~(\ref{current_phase_BdG_analytical}) suggests quite complex behavior of the spontaneous phase difference within the parameter range $h<\mu_f$. It is clear that in the general case the anomalous phase is determined by the competition of all the above contributions. The important point in the subsequent analysis is that these contributions are of different magnitude and exhibit different behavior with respect to the superconducting phase difference $\varphi$. In particular, the first two terms proportional to $\sin(\varphi)$ in the right-hand side of Eq.~(\ref{current_phase_BdG}) favor the spontaneous phase $\psi = 0$ or $\pi$ whereas the remaining terms are responsible for the tunability of $\psi$ as a function of $\chi$. Corresponding critical currents $I_{\uparrow\downarrow}^{\pm}$, $I_{\uparrow\uparrow}$ and $I_{\downarrow\downarrow}$ can, in turn, exhibit a singular behavior when $p_{\uparrow,\downarrow}L = \pi n$ ($n = 1,2,3,...$). One can naturally expect that such jumpwise behavior of the critical currents should result in the jumpwise changes and/or oscillations of the spontaneous superconducting phase difference $\psi$ upon the change in the band structure parameters $\mu_f$ and $h$ of the central ferromagnet as well as the junction length $L$. It is important to note that even though the validity of Eqs.~(\ref{current_phase_BdG_analytical}) breaks down near the resonances, our analytical results also provide qualitative explanation for $\psi(h)$ curves observed in our numerical simulations.

\subsection{Numerical simulations}

We proceed with the discussion of the results of numerical simulations. Our focus is on the behavior of the anomalous phase $\psi$ as a function of the exchange field in the central ferromagnet $h$ and the misorientation angle $\chi$. For simplicity, we consider the case of equal barrier strength parameters $Z_{0,1} = Z_{0,2} = Z_0$, $Z_1 = Z_2 = Z$ and take $\theta_1 = \theta_2 = \pi/2$. Our numerical results are obtained for the following parameter set: $T = 0.1\Delta_0$, $\mu_s/\Delta_0 = 10^3$, $\mu_f/\Delta_0 = 5\times 10^2$, $m_s = m_f$, $Z_0 = 0$, and $Z = 0.5$. 

 \begin{figure*}[htpb]
\centering
\includegraphics[scale = 0.8]{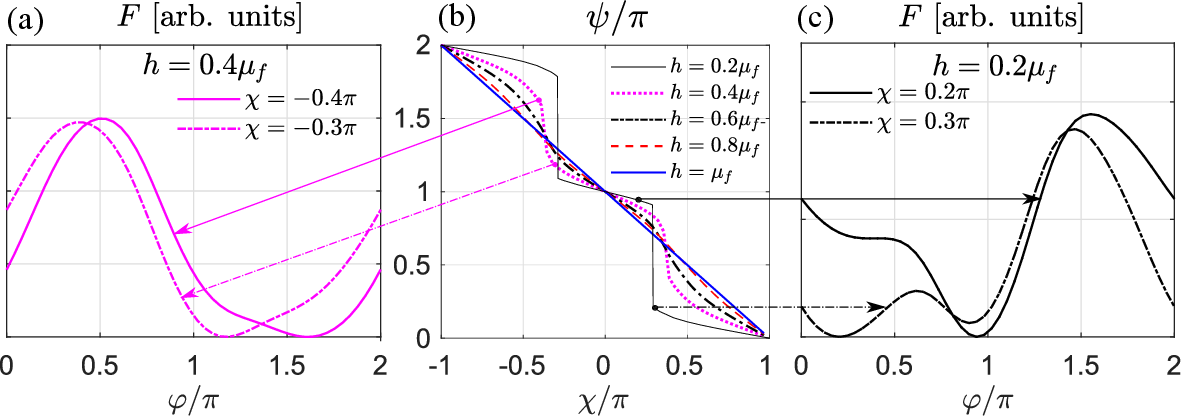}
\caption{(b) Typical plots of the ground-state superconducting phase difference $\psi$ versus the misorientation angle $\chi$ for $h/\mu_f = 1$, $0.8$, $0.6$, $0.4$, and $0.2$. Corresponding $F(\varphi)$ dependencies for $h/\mu_f = 0.4$, $\chi/\pi = -0.4$, $-0.3$ and $h = 0.2\mu_f$, $\chi/\pi = 0.2$, $0.3$ are shown in (a) and (c), respectively. We choose $L = 0.02\xi$ to produce the plots.
}
\label{Fig:anomalous_phase_versus_chi}
\end{figure*}

 \begin{figure*}[htpb]
\centering
\includegraphics[scale = 0.63]{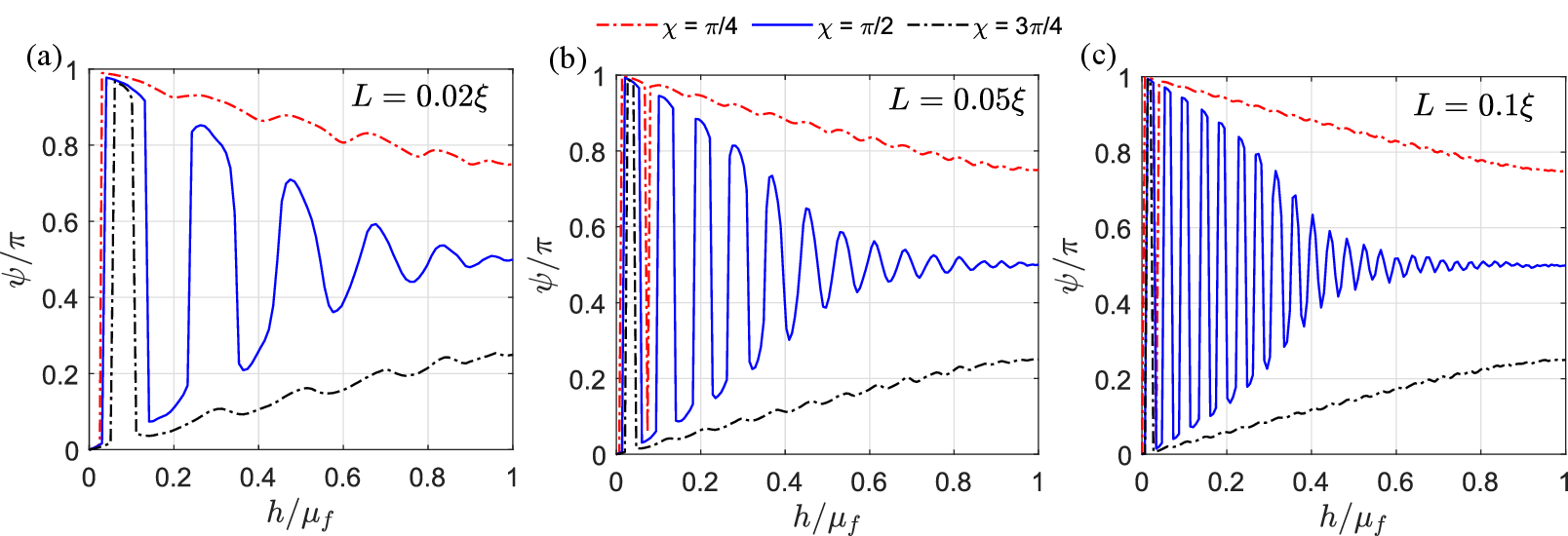}
\caption{Typical plots of the ground-state $\psi(h)$ dependencies within the full range of the exchange fields for $\chi = \pi/4$, $\pi/2$, and $3\pi/4$. Panels (a), (b), and (c) correspond to $L/\xi = 0.02$, 0.05, and 0.1, respectively. 
}
\label{Fig:anomalous_phase_versus_h}
\end{figure*}

Typical current-phase relations $I_s(\varphi)$ along with the corresponding $F(\varphi)$ curves for $\chi = 0$, $\pi/4$, $\pi/2$, $3\pi/4$, $\pi$ and several exchange fields $h/\mu_f = 1$, $0.8$, and $0.6$ are shown in Fig~\ref{Fig:results_BdG}. Circles in Fig.~\ref{Fig:results_BdG} denote the superconducting phase difference, at which the free energy of the junction reaches its minimal value (the anomalous phase $\psi$). 
In the half-metallic regime $h/\mu_f = 1$ [see Figs.~\ref{Fig:results_BdG}(a) and~\ref{Fig:results_BdG}(b)] the corresponding $I_s(\varphi)$ curves are sinusoidal and the spontaneous phase $\psi \approx \pi - \chi $. The decrease in $h$ [see panels (c)-(f) in Fig.~\ref{Fig:results_BdG}] leads to deviations of the spontaneous phase from $\pi - \chi$ and the appearance of the superconducting diode effect in the system $I_{c+}\neq I_{c-}$, where $I_{c+} = \max_{\varphi} I_s(\varphi)$ and $I_{c-} = |\min_{\varphi}I_s(\varphi)|$. Note that a nonreciprocal superconducting transport is a common feature of the considered Josephon junctions with non-coplanar magnetization distribution and $h < \mu_f$~\cite{Margaris,Halterman}. The results in Figs.~\ref{Fig:results_BdG}(c) and~\ref{Fig:results_BdG}(e) clearly demonstrate that both the magnitude of the diode effect as well as the preferential direction, for which the Josephson junction can carry larger supercurrent, are governed by the misorientation angle $\chi$.

Let us now discuss the tunability of the ground-state superconducting phase difference 
as a function of the hybrid structure parameters. We show several $\psi(\chi)$ plots for $h/\mu_f = 1$, $0.8$, $0.6$, $0.4$ and $0.2$ in Fig.~\ref{Fig:anomalous_phase_versus_chi}(b). One can see that in the half-metallic regime for $h/\mu_f = 1$, the corresponding $\psi(\chi)$ dependence is, indeed, linear $\psi = \pi - \chi$. As it has been explained previously, such behavior originates from the fact that the supercurrent is carried only by the parallel spin-triplet Cooper pairs from the lower spin-split subband in the central ferromagnet. It is interesting to note that the tunability of the anomalous phase can persist up to rather small values of the exchange field $h$. Moreover, Fig.~\ref{Fig:anomalous_phase_versus_chi}(a) also demonstrates that within the 
parameter range $h < \mu_f$, the spontaneous phase difference can exhibit sudden jumps at certain $\chi$ values due to the competition between two local minima of the free energy of the contact versus the superconducting phase difference [see Figs.~\ref{Fig:anomalous_phase_versus_chi}(a) and~\ref{Fig:anomalous_phase_versus_chi}(c)]. The results demonstrate that for the considered Josephson junctions with non-coplanar magnetic texture, the variations of the misorientation angle in the case $h<\mu_f$ can induce the first-order phase transitions between the states with different anomalous phases accompanied with the hysteresis phenomena.

 \begin{figure}[htpb]
\centering
\includegraphics[scale = 0.18]{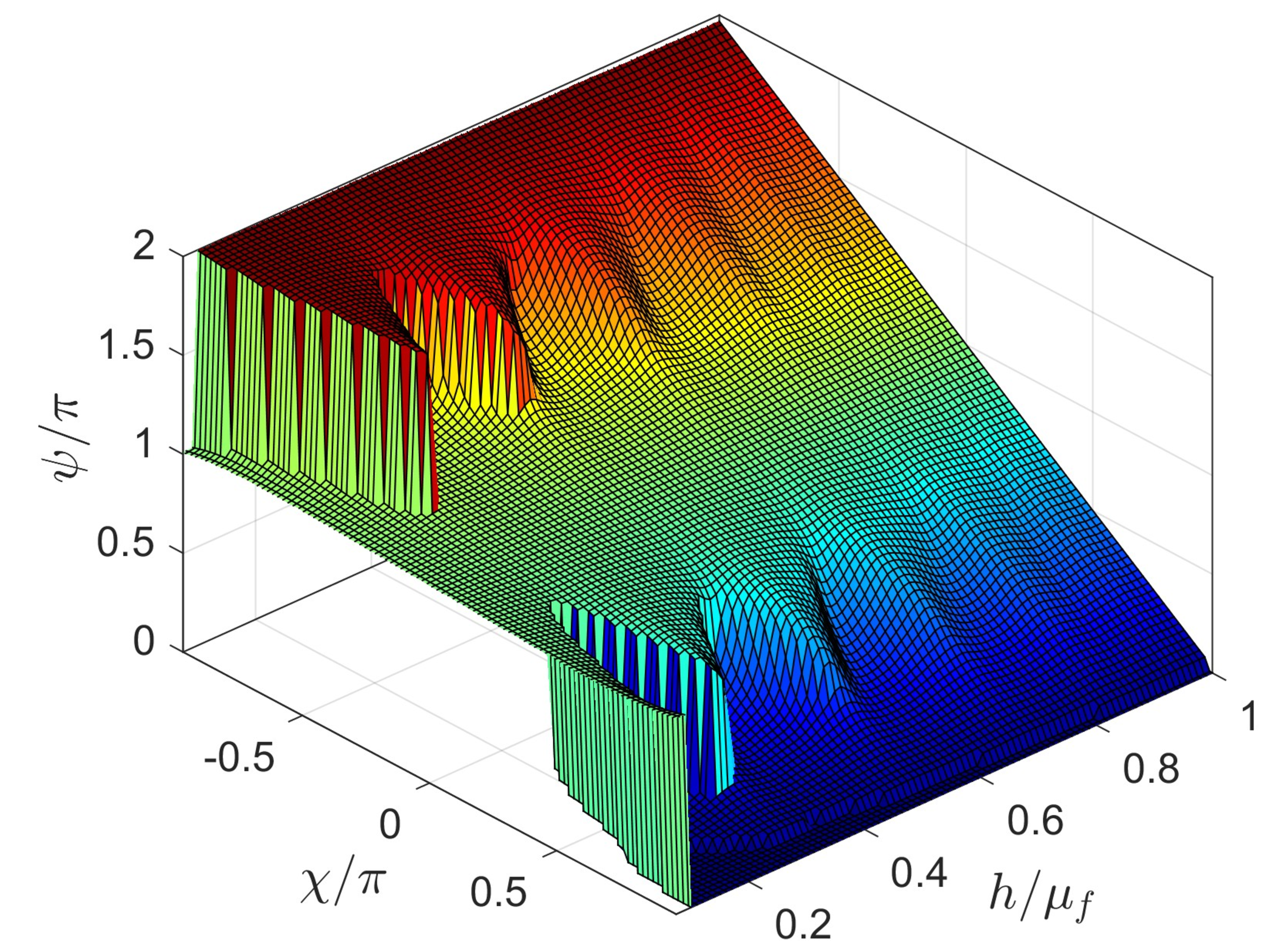}
\caption{Surface plot of the ground-state superconducting phase difference $\psi$ versus $\chi$ and $h$ for $L = 0.02\xi_s$.}
\label{Fig:anomalous_phase_versus_chi_and_h}
\end{figure}

\begin{figure*}[htpb]
	\includegraphics[scale=0.7]{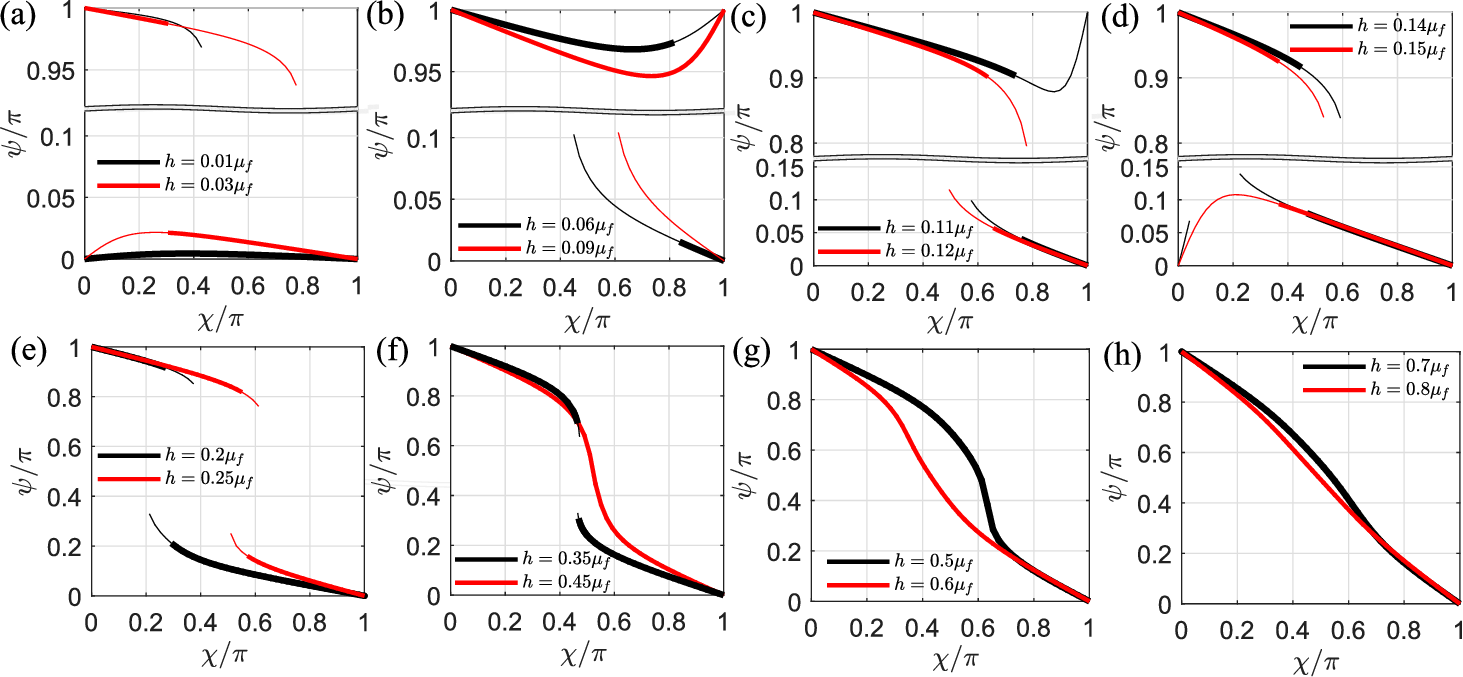}
	\caption{Typical dependencies of the superconducting phase difference $\psi$ on the misorientation angle $\chi$ for several exchange fields $h$ in the central ferromagnet. Thick (thin) lines show the results corresponding to the global (local) minimum of the junction free energy. The results for $\chi\in[-\pi, 0]$ are obtained via inversion relative to the point $(\psi,\chi) = (\pi, 0)$. We take $h/\mu_f = 0.01$ and 0.03 in (a), 0.06 and 0.09 in (b), 0.11 and 0.12 in (c), 0.14 and 0.15 in (d), 0.2 and 0.25 in (e), 0.35 and 0.45 in (f), 0.5 and 0.6 in (g), 0.7 and 0.8 in (h). We choose $L = 0.02\xi$ to produce the plots.}  
	\label{Fig:pumping}
\end{figure*}

Typical behavior of the ground-state superconducting phase difference 
within the full range of the exchange fields for $\chi = \pi/4$, $\pi/2$, and $3\pi/4$ are shown in Fig.~\ref{Fig:anomalous_phase_versus_h}. The results in Figs.~\ref{Fig:anomalous_phase_versus_h}(a), \ref{Fig:anomalous_phase_versus_h}(b), and \ref{Fig:anomalous_phase_versus_h}(c) are obtained for various length of the junction $L/\xi = 0.02$, 0.05, and 0.1, respectively. Let us, first, discuss the plots in Fig.~\ref{Fig:anomalous_phase_versus_h}(a). One can see that for $\chi = \pi/4$ the spontaneous phase jumps at $h \approx 0.05\mu_f$ and then exhibits several oscillations upon the increase in the exchange field. At $h \approx \mu_f$ the anomalous phase saturates at $\pi - \chi$. Except the case of rather weak exchange fields, the behavior of the spontaneous phase for $\chi = 3\pi/4$ is qualitatively similar to the above discussed one. For $\chi = \pi/2$ we observe that the spontaneous phase exhibits both oscillations and the jumps. It is important to note that the number of oscillations on $\psi(h)$ curves in Fig.~\ref{Fig:anomalous_phase_versus_h}(a) is of order $N_f = k_fL/\pi$, where $k_f$ is the Fermi momentum in the central layer at $h = 0$. Indeed, for our choice of the hybrid structure parameters $N_f \sim 10$, which allows us to associate the oscillations of the anomalous phase with the size quantization effects in the central F layer. The results for longer junctions $L/\xi = 0.05$ and 0.1 [Figs.~\ref{Fig:anomalous_phase_versus_h}(b) and \ref{Fig:anomalous_phase_versus_h}(c)] reveal the increase in the number of oscillations upon the increase in the junction length. One can see that the number of oscillations of the anomalous phase in Figs.~\ref{Fig:anomalous_phase_versus_h}(b) and \ref{Fig:anomalous_phase_versus_h}(c) is of order $k_fL/\pi$. The resulting exchange-field behavior of the anomalous phase is in a good qualitative agreement with the results of Eq.~(\ref{current_phase_BdG_analytical}). Our main results regarding the behavior of the ground-state superconducting phase difference are summarized in Fig.~\ref{Fig:anomalous_phase_versus_chi_and_h}, where we show the
surface plot of the anomalous phase $\psi$ versus the exchange field $h$ and the misorientation angle $\chi$. Qualitatively, the observed behavior of the anomalous phase reflects the interplay between several physical phenomena such as the spin-filtering effect, the size quantization effect in ballistic junctions with rather thin ferromagnet and a tunability of the spin-triplet supercurrent by the relative orientation of the magnetic moments in three ferromagnetic layers.  

Finally, we analyze the features of the phase pumping phenomenon for different exchange fields $h$ in the central ferromagnet. For this purpose we perform the calculations of the superconducting phase difference corresponding to both global and local minima of the free energy. The results presented in Fig.~\ref{Fig:pumping} show the detailed evolution in the behavior of the system upon the increase in $h$. In particular, the plots for $h = 0.01\mu_f$ (shown by black lines in Fig.~\ref{Fig:pumping}(a)) reveal two states of the system, one of which is stable within the whole range of misorientation angles. The other one is metastable and can be realized only within a certain $\chi$ range. So, in this case the mutual rotation of magnetization in F$_1$ and F$_2$ layers corresponds to a topologically trivial trajectory of the system in the parameter space $(\psi,\chi)$ (see, e.g., the black solid line in Fig.~\ref{Fig:torus_figure}) with the superconducting phase difference trapped near zero. For $h = 0.03\mu_f$ (see the red lines in Fig.~\ref{Fig:pumping}(a)) there appear stable states both near 0 and $\pi$ within certain $\chi$ ranges. Thus, the topologically trivial evolution of the superconducting phase upon the magnetization rotation can be accompanied with thermally activated jumps of the spontaneous phase. The probability of the latter process obviously decreases with decreasing temperature. Note that for $h/\mu_f = 0.06$, 0.09, and 0.11 (see Fig.~\ref{Fig:pumping}(b) and the black lines in Fig.~\ref{Fig:pumping}(c)) the system exhibits a similar behavior except that the superconducting phase difference can be trapped near $\pi$. Corresponding plots for $h/\mu_f = 0.12$ (see the red lines in Fig.~\ref{Fig:pumping}(c)) reveal the change in the topology of the phase evolution and the appearance of the phase pumping: the gain of the Josephson phase at the pumping period switches to $2\pi$. The presence of the jump of the ground-state phase difference and the metastable states in the vicinity of the jump implies that the nontrivial phase evolution should be accompanied with the hysteresis phenomena. The results for $h/\mu_f = 0.15$ (shown by the red lines in Fig.~\ref{Fig:pumping}(c)) illustrate the transition into the topologically trivial state. We observe another transition into the state with a nontrivial phase evolution at $h \approx 0.2\mu_f$ (see the black lines in Fig.~\ref{Fig:pumping}(e)). Figs.~\ref{Fig:pumping}(f)-(h) demonstrate that a nontrivial gain of the superconducting phase at the rotation period persists upon further increase in $h$, the $\chi$ regions corresponding to metastable states shrink and disappear for rather large $h$ (see, e.g., the results for $h/\mu_f = 0.45$ in Fig.~\ref{Fig:pumping}(f)). Therefore, we find that for rather large exchange fields in the central ferromagnet the magnetization rotation corresponds to a continuous topologically nontrivial trajectory of the system in the parameter space $(\psi,\chi)$ (see, e.g., the red solid line in Fig.~\ref{Fig:torus_figure}) and the continuous phase pumping whereas for smaller fields the corresponding trajectories are discontinuous.

\section{Conclusion}\label{Sec_conc}

To sum up, we have studied the anomalous Josephson effect in S/F$_1$/F/F$_2$/S systems with non-coplanar magnetic moments in ferromagnetic layers with arbitrary ratio between the exchange field in the F layer and the Fermi energy. 

As a first step, considering SF$_1$/half-metal/SF$_2$ Josephson junctions of atomic thickness in the frames of Gor'kov formalism we have demonstrate that the spontaneous ground-state phase difference $\psi$ arising in such systems coincides with the angle between projections of magnetic moments in the SF$_1$ and SF$_2$ layers to the plane perpendicular to the spin quantization axis in half-metal. Interestingly, this feature of $\psi$ junctions gives rise to the Berry phase effects. Indeed, the rotation of magnetic moment, e.g., in the SF$_1$ layer should produce the accumulation of the ground-state phase $2\pi$ after the full precession period which may results in the interesting possibility of magnetic flux pumping in superconducting loop \cite{Meng_Flux}. Also, using the exact solution of Gor'kov equations for the Green functions we have calculated the critical current in the current-phase relation and establish its dependence on the magnetic moments direction in SF$_1$ and SF$_2$ layers.

Then we analyzed the behavior of the ground-state superconducting phase difference and the features of the phase pumping effect in S/F$_1$/F/F$_2$/S junctions upon the decrease in the exchange field in the F layer from the values comparable with the Fermi energy (the limit of half-metal) down to the ones of the order of the critical temperature of the superconducting phase transition. To do this we have performed calculations of the Josephson transport for clean S/F$_1$/F/F$_2$/S systems with the finite F layer thickness. For a half-metallic central layer it has been shown that the spontaneous phase is a linear function of the misorientation angle $\psi = \pi - \chi$. As the exchange field decreases we have found that the considered Josephson systems are characterized by a nonlinear $\psi(\chi)$ dependence and can feature two competing local minima in the free energy of the junction vs the superconducting phase difference. This competition leads to a bistable behavior of the system, which manifests itself through the jump-wise changes of the spontaneous phase upon the variation in the misorientation angle accompanied with the hysteresis phenomena. We have demonstrated that further decrease in $h$ can induce several changes in the topology of the phase evolution: the gain of the Josephson phase at the rotation period switches between 0 to $2\pi$. Also we have found that the spontaneous phase can exhibit oscillating and/or the jump-wise behavior as a function of the exchange field. 

The $\psi$ junctions incorporated into superconducting circuits provide a possibility to switch between the superconducting states with different vorticities without the application of external magnetic field \cite{Meng_Flux}. The direct coupling between superconducting phase difference and orientation of the magnetic moment open a way to generate a magnetic moment precession by superconducting current in $\psi$ junction similar to that predicted for $\varphi_0$ junction \cite{Shukrinov}. We believe that the $\psi$ junction may be a very interesting building block for superconducting spintronics.

Now there are several experimental evidences of the Josephson current  through the half- metal ferromagnets \cite{Keizer, Anwar, Visani, Sanchez-Manzano, Jungxiang}. Most probably, this long-ranged triplet supercurrent could be generated by the non-collinear surface magnetization at the interface with a half metal. In such a  case, our model of Sec.~\ref{Sec_BdG} with the spin-active interfaces seems to be quite adequate for the description of the experiments \cite{Keizer, Anwar, Visani, Sanchez-Manzano, Jungxiang}. The non-collinear surface magnetization should be the same at both ends of the junctions and then we should have $\psi=\pi$ , i.e. the $\pi$ junction realization. This may be directly verified by incorporating Josephson junction with a half-metal in a closed superconducting loop (similar to the experiments \cite{Bauer}). Note that recently a superconducting quantum interference device was used to detect the transitions between $0$ and $\pi$ states in S/F/S junctions with composite ferromagnetic layer \cite{Birge_1, Birge_2}. This experimental technique and the fabricated setups provide a perfect playground to verify the effects predicted in the present paper and we hope that our results will stimulate the corresponding activity.

\section*{Acknowledgements}

The authors thank A. A. Bespalov for stimulating discussions. This work was supported by Ministry of Science and Higher Education of
the Russian Federation within the framework of state funding for the creation
and development of World-Class Research Center (WCRC) ``Digital
biodesign and personalized healthcare,'' grant no. 075-15-2022-304 in part related to the analysis of the system of atomically thin layers and the Russian Science Foundation (Grant No. 20-12-00053) in part related to the analysis of the current-phase relations for the system with finite thickness of the central ferromagnetic layer. The work of A. I. B. was supported by ANR SUPERFAST and the LIGHT S\&T Graduate Program and EU COST CA21144  Superqumap. The work of H. M. was supported by the National Natural Science Foundation of China (Grant No. 12174238), the Natural Science Basic Research Program of Shaanxi (Program No. 2020JM-597), and the Scientific Research Foundation of Shaanxi University of Technology (Grant No. SLGKY2006).

\appendix
\section{Critical temperature of SF/HM superlattice}\label{AppendixA}

Assuming $t$ to be small, we perform the power expansion of (\ref{AGF_Tc}) up to the 4th order:
\begin{multline}
\frac{\hat F^{\dagger}}{\Delta_0}=X_{+} \hat I X_{-}+|T(k+q)|^2 X_{+} \hat I X_{-} Y_{-} X_{-} \\+|T(q)|^2 X_{+}Y_{+} X_{+} \hat I X_{-} +|T(q)|^4 X_{+}Y_{+} X_{+} Y_{+} X_{+} \hat I X_{-} +\\+ |T(k+q)|^4 X_{+} \hat I X_{-} Y_{-} X_{-} Y_{-} X_{-} \\+ |T(q)|^2 |T(k+q)|^2 X_{+}Y_{+} X_{+} \hat I X_{-} Y_{-} X_{-},
\end{multline}
where $X_{\pm}=(i\omega_n \pm \hat C)^{-1}$, $Y_{\pm}=(i\omega_n \pm \hat P)^{-1}$. Then we calculate the following expressions entering Eq.(\ref{SCE_Tc_general}):
\begin{equation}
\sum_{n=-\infty}^{\infty} \int^{+\infty}_{-\infty} d \xi \int^{\pi}_{-\pi}\frac{dq}{2\pi} X_{+} \hat I X_{-}=-\sum_{n>0} \frac{2\pi \omega_n}{\omega_n^2+h^2} \hat I,
\end{equation}
\begin{multline}
\sum_{n=-\infty}^{\infty} \int^{+\infty}_{-\infty} d \xi \int^{\pi}_{-\pi}\frac{dq}{2\pi} [|T(k+q)|^2 X_{+} \hat I X_{-} Y_{-} X_{-} \\+|T(q)|^2 X_{+}Y_{+} X_{+} I X_{-}]=\sum_{n>0} \frac{4\pi t^2 \omega_n (\omega_n^2-2h^2)}{(\omega_n^2+h^2)^2 (4\omega_n^2+h^2)}\hat I,
\end{multline}
\begin{multline}
\sum_{n=-\infty}^{\infty} \int^{+\infty}_{-\infty} d \xi \int^{\pi}_{-\pi}\frac{dq}{2\pi} \left[|T(q)|^4 X_{+}Y_{+} X_{+} Y_{+} X_{+} \hat I X_{-} \right.\\\left.+ |T(k+q)|^4 X_{+} \hat I X_{-} Y_{-} X_{-} Y_{-} X_{-}\right]=\\=-\frac{3\pi t^4}{8}\sum_{n>0}\left[ \frac{4(h^4-9h^2 \omega_n^2+2\omega_n^4)}{\omega_n(\omega_n^2+h^2)^3 (4\omega_n^2+h^2)}\right.\\ \left.- \frac{2h^2(h^6+7h^4 \omega_n^2-46 h^2 \omega_n^4 -16 \omega_n^6)}{\omega_n^3(\omega_n^2+h^2)^3 (4\omega_n^2+h^2)^2}\cos^2 \theta\right]\hat I,
\end{multline}
\begin{multline}
\sum_{n=-\infty}^{\infty} \int^{+\infty}_{-\infty} d \xi \int^{\pi}_{-\pi}\frac{dq}{2\pi} \left[|T(q)|^2 |T(k+q)|^2 X_{+}Y_{+} X_{+} \right.\\ \left.+ \hat I X_{-} Y_{-} X_{-}\right]=\\ =t^4\left(1+\frac{\cos k}{2} \right)\sum_{n>0}\frac{\pi(h^4+35h^2 \omega_n^2+70\omega_n^4)}{\omega_n(\omega_n^2+h^2)^3 (4\omega_n^2+h^2)^2}\sin^2 \theta \hat I.
\end{multline}
We substitute the above expressions into the self-consistency equation (\ref{SCE_Tc_general}) and obtain:
\begin{multline}
\label{SCE_Tc}
{\rm ln}\left(\frac{T_{c}}{T_{c0}} \right)=-2\pi T_c \sum_{n>0} \left[\frac{h^2}{\omega_n(\omega_n^2+h^2)}\right. \\+\frac{2t^2\omega_n (\omega_n^2-2h^2)}{(\omega_n^2+h^2)^2 (4\omega_n^2+h^2)} \\- \frac{3t^4(h^2-2\omega_n^2)^2(4\omega_n^4-9h^2 \omega_n^2-h^4)}{8\omega_n^3(\omega_n^2+h^2)^3 (4\omega_n^2+h^2)^2}+\\+\frac{t^4h^2(328 \omega_n^6+278h^2 \omega_n^4-17 h^4 \omega_n^2 -3 h^6)}{8\omega_n^3(\omega_n^2+h^2)^3 (4\omega_n^2+h^2)^3}\sin^2 \theta \\ \left.+\frac{t^4h^2(h^4+35h^2 \omega_n^2 +70 \omega_n^4)\cos k}{4\omega_n(\omega_n^2+h^2)^3 (4\omega_n^2+h^2)^2}\sin^2 \theta \right].
\end{multline}

Finally, representing the critical temperature in the form
\begin{equation}
\label{App_Tc}
T_{c}={\tilde T}_{c0}(1+a t^2+b t^4),
\end{equation}
we calculate $a$ and $b$ using (\ref{SCE_Tc}) and obtain
\begin{equation}
a=-\frac{1}{W}\sum_{n>0}\frac{4\pi {\tilde T}_{c0} \tilde\omega_n(\tilde\omega_n^2-2h^2)}{(\tilde\omega_n^2+h^2)^2 (4\tilde\omega_n^2+h^2)},
\end{equation}
\begin{multline}
\label{App_b}
b=\frac{2\pi {\tilde T}_{c0}}{W}\sum_{n>0} \left[\frac{3(h^2-2\tilde\omega_n^2)^2(4\tilde\omega_n^4-9h^2 \tilde\omega_n^2-h^4)}{8\tilde\omega_n^3(\tilde\omega_n^2+h^2)^3 (4\tilde\omega_n^2+h^2)^2}\right. \\-\frac{h^2(328 \tilde\omega_n^6+278h^2 \tilde\omega_n^4-17 h^4 \tilde\omega_n^2 -3 h^6)}{8\tilde\omega_n^3(\tilde\omega_n^2+h^2)^3 (4\tilde\omega_n^2+h^2)^3}\sin^2 \theta-\\ \left.-\frac{h^2(h^4+35h^2 \tilde\omega_n^2 +70 \tilde\omega_n^4)\cos k}{4\tilde\omega_n(\tilde\omega_n^2+h^2)^3 (4\tilde\omega_n^2+h^2)^2}\sin^2 \theta \right]\\+\frac{a^2}{2W} \left[1-\sum_{n>0}\frac{4\pi {\tilde T}_{c0} h^2 \tilde\omega_n(3\tilde\omega_n^2+h^2)}{(\tilde\omega_n^2+h^2)^3} \right]+\\+\frac{a}{W}\sum_{n>0}\frac{16\pi {\tilde T}_{c0} \tilde\omega_n(2\tilde\omega_n^6 -10 \tilde\omega_n^4 h^2-2\tilde\omega_n^2 h^4+h^6)}{(\tilde\omega_n^2+h^2)^3 (4\tilde\omega_n^2+h^2)^2}.
\end{multline}
where $\tilde\omega_n=\pi\tilde T_{c0}(2n+1)$. Since $h \ll \tilde\omega_n$, $W>0$ and the critical temperature is higher for the $\pi$-phase.

\begin{widetext}
\section{Details of numerical calculations}\label{AppendixB}

Here we provide the details of numerical simulations. Our starting point is the expression for the free energy~(\ref{BdG_junction_energy}). The determinant of the matching condition matrix $\mathcal{P}$ is defined by Eq.~(\ref{determinant_solvability_condition}) and the scattering matrices $\mathcal{K}_i$ ($i = 1,2$) are given in Eq.~(\ref{scattering_matrices}). As a first step, we calculate the matrix product $\check{\mathcal{K}}_1(\omega_n,\mathbf{p}_{||})\check{\mathcal{K}}_2(\omega_n,\mathbf{p}_{||})$ for a certain Matsubara frequency and the in-plane momentum:
\begin{eqnarray}\label{matrix_product_appendix}
 \check{\mathcal{K}}_1\check{\mathcal{K}}_2 = \check{Q}(L/2)(\check{W}_1 + \check{K})^{-1}(\check{W}_1 -\check{K})\check{Q}(L)(\check{W}_2 + \check{K})^{-1}(\check{W}_2 - \check{K})\check{Q}(L/2) \ .
\end{eqnarray}
For the case $\theta_1 = \theta_2 = \pi/2$ considered in the main text we have $\mathbf{n}_j\hat{\boldsymbol{\sigma}} = \cos(\chi_j)\hat{\sigma}_x + \sin(\chi_j)\hat{\sigma}_y$, and the matrix products can be reduced to the form
\begin{eqnarray}\label{matrix_products_intermediate}
 (\check{W}_j + \check{K})^{-1}(\check{W}_j - \check{K}) = \check{\Lambda}_j 
 [(g + iZ_j\hat{\sigma}_x)\check{\tau}_z + iZ_{0,j} + \check{K} - f\check{\tau}_y]^{-1}
 [(g + iZ_j\hat{\sigma}_x)\check{\tau}_z + iZ_{0,j} - \check{K} - f\check{\tau}_y]\check{\Lambda}_j^{\dagger} \ .
\end{eqnarray}
Here we introduced the unitary matrices $\check{\Lambda}_j = e^{i(\varphi_j\check{\tau}_z - \chi_j\hat{\sigma}_z)/2}$, the values $\varphi_1$ and $\varphi_2$ stand for the phase of the superconducting order parameter in the left and right lead, respectively. We find the inverse matrix entering Eq.~(\ref{matrix_products_intermediate}) analytically and then derive the expression for the matrix product
\begin{eqnarray}\label{matrix_product_explicit}
 \check{\Lambda}_j^{\dagger}(\check{W}_j + \check{K})^{-1}(\check{W}_j - \check{K})\check{\Lambda}_j = 
 \begin{bmatrix}1 - 2\hat{\mathcal{W}}_j^{\rm T}(iZ_{0,j}-g-iZ_j\hat{\sigma}_x + \hat{\bar{K}})\hat{K}&2if\hat{\mathcal{W}}_j^{\rm T}\hat{\bar{K}}\\ -2if\hat{\mathcal{W}}_j\hat{K}&1 - 2\hat{\mathcal{W}}_j(iZ_{0,j}+g + iZ\hat{\sigma}_x + \hat{K})\hat{\bar{K}}\end{bmatrix} \ ,
\end{eqnarray}
where 
\begin{subequations}\label{various_matrices_appendix}
 \begin{align}
  \hat{\mathcal{W}}_j = -\frac{1}{\eta_j}\begin{bmatrix}w_j - iZ_{0,j}s_{\downarrow\downarrow} + g\delta_{\downarrow\downarrow}-K_{\downarrow}\bar{K}_{\downarrow}&-iZ_j(2g + \delta_{\uparrow\downarrow})\\-iZ_j(2g + \delta_{\downarrow\uparrow})&w - iZ_{0,j}s_{\uparrow\uparrow}+g\delta_{\uparrow\uparrow}-K_{\uparrow}\bar{K}_{\uparrow}\end{bmatrix} \ ,\\
  \eta_j = [w_j - iZ_{0,j}s_{\uparrow\uparrow}+g\delta_{\uparrow\uparrow}-K_{\uparrow}\bar{K}_{\uparrow}][w-iZ_{0,j}s_{\downarrow\downarrow}+g\delta_{\downarrow\downarrow}-K_{\downarrow}\bar{K}_{\downarrow}] + Z_j^2(2g + \delta_{\uparrow\downarrow})(2g + \delta_{\downarrow\uparrow}) \ ,
 \end{align}
\end{subequations}
$w_j = 1 + Z_0^2 - Z^2$, $s_{\sigma\sigma^\prime} = K_{\sigma} + \bar{K}_{\sigma^\prime}$, $\delta_{\sigma\sigma^\prime} = K_{\sigma}-\bar{K}_{\sigma^\prime}$, and $\sigma,\sigma^\prime = \uparrow,\downarrow$ are the spin indices. The determinant of the solvability condition matrix $\mathcal{P}(i\omega_n,\mathbf{p}_{||})$ [see Eq.~(\ref{determinant_solvability_condition})] is calculated numerically using the above expressions~(\ref{matrix_product_appendix}), (\ref{matrix_product_explicit}), and (\ref{various_matrices_appendix}). The next steps are the integration of $\log|\mathcal{P}(i\omega_n,\mathbf{p}_{||})|$ with respect to the parallel momentum for a certain Matsubara frequency and then the summation over the Matsubara frequencies. The dominant contribution to the momentum integrals stems from the states with $|\mathbf{p}_{||}|<p_{Fs}$, where $p_{Fs}$ is the Fermi momentum in the normal-metal state of the superconducting lead. Note that some care is needed for numerical evaluation of the wave numbers $p_{\uparrow,\downarrow}(i\omega_n)$, $\bar{p}_{\uparrow,\downarrow}(i\omega_n)$ defined by Eq.~(\ref{wave_numbers_ferromagnet}). In Eqs.~(\ref{current_phase_BdG_definition}) and (\ref{BdG_junction_energy}) the summation is carried out over positive Matsubara frequencies. In our numerical calculations we choose the branch cut of the complex square root to be along the negative real axis. For this particular choice one should couple the electronic states with the wave numbers $p_{\sigma}$ in the ferromagnet with the hole states with the wave numbers $-\bar{p}_{\sigma}$ [see Eqs.~(\ref{ferromagnet_solution}) and (\ref{ferromagnet_parameters})] because in the opposite case the kinematic phase factors of the hole excitations $e^{i\bar{p}_{\sigma}L}$ quickly diverge upon the increase in the Matsubara frequency. This problem is resolved by the replacement $\bar{p}_{\sigma}\to - \bar{p}_{\sigma}$ in numerical code. 
As a result, we obtain the free energy of the junction $F$~(\ref{BdG_junction_energy}) as a function of the superconducting phase difference $\varphi$. Finally, we compute the derivative $\partial F/\partial\varphi$, which gives us the current-phase relation~(\ref{current_phase_BdG_definition}). 
\end{widetext}


\begin{thebibliography}{9}

\bibitem{AltshulerS1999} B.~L. Altshuler and L.~I. Glazman, Science \textbf{283}, 1864 (1999).
\bibitem{SwitkesS1999} M. Switkes, C.~M. Marcus, K. Campman, and A.~C. Gossard, Science \textbf{283}, 1905 (1999).
\bibitem{MoskaletsPRB2005} M. Moskalets and M. B\"{u}ttiker, Phys. Rev. B \textbf{72}, 035324 (2005).
\bibitem{ButtikerJLTP2000} M. B\"{u}ttiker, J. Low Temp. Phys. \textbf{118}, 519 (2000).

\bibitem{ThoulessPRB1983} D.~J.~Thouless, Phys. Rev. B \textbf{27}, 6083 (1983).

\bibitem{Nazarov_PRL} V. Braude, and Yu. V. Nazarov, Phys. Rev. Lett. \textbf{98}, 077003 (2007).

\bibitem{AstafievN2022} R.~S. Shaikhaidarov, K.~H. Kim, J.~W. Dunstan, I.~V. Antonov, S. Linzen, M. Ziegler, D.~S. Goluber, V.~N. Antonov, E.~V. Il'ichev, and O.~V. Astafiev, Nature \textbf{608}, 45 (2022). 
	

\bibitem{Buzdin} A. Buzdin, Phys. Rev. Lett. \textbf{101}, 107005 (2008).

\bibitem{Ustinov} A. V. Ustinov, V. K. Kaplunenko, J. Appl. Phys. \textbf{94}, 5405 (2003).

\bibitem{Bauer} A. Bauer, J. Bentner, M. Aprili, M. L. Della-Rocca, M. Reinwald, W. Wegscheider, C. Strunk, Phys. Rev. Lett. \textbf{92}, 217001 (2004).

\bibitem{Buzdin_2005} A. Buzdin, Phys. Rev. B \textbf{72}, 100501(R) (2005).

\bibitem{Feofanov} A. K. Feofanov, V. A. Oboznov, V. V. Bol'ginov, J. Lisenfeld, S. Poletto, V. V. Ryazanov, A. N. Rossolenko, M. Khabipov, D. Balashov, A. B. Zorin, P. N. Dmitriev, V. P. Koshelets, A. V. Ustinov, Nat. Phys. \textbf{6}, 593 (2010).
 
\bibitem{Ortlepp} T. Ortlepp, Ariando, O. Mielke, C. J. M. Verwijs, K. F. K. Foo, H. Rogalla, F. H. Uhlmann, H. Hilgenkamp, Science \textbf{312}, 1495 (2006).

\bibitem{Geshkenbein} V. B. Geshkenbein and A. I. Larkin, Pis’ma Zh. Eksp. Teor. Fiz. \textbf{43}, 306 (1986) [JETP Lett. \textbf{43}, 395 (1986)].

\bibitem{Yip} S. Yip, Phys. Rev. B \textbf{52}, 3087 (1995).

\bibitem{Sigrist} M. Sigrist, Prog. Theor. Phys. \textbf{99}, 899 (1998).

\bibitem{Kashiwaya} S. Kashiwaya, and Y. Tanaka, Rep. Prog. Phys. \textbf{63}, 1641
(2000).

\bibitem{Tanaka_TI} Y. Tanaka, T. Yokoyama, N. Nagaosa, Phys. Rev. Lett. \textbf{103}, 107002 (2009).

\bibitem{Houzet_TI} F. Dolcini, M. Houzet, J. S. Meyer, Phys. Rev. B \textbf{92}, 035428 (2015).

\bibitem{Aubin} A. Assouline, C. Feuillet-Palma, N. Bergeal, T. Zhang, A. Mottaghizadeh, A. Zimmers, E. Lhuillier, M. Eddrie, P. Atkinson, M. Aprili, H. Aubin, Nat. Commun. \textbf{10}, 126 (2019).

\bibitem{Reynoso} A. A. Reynoso, G. Usaj, C. A. Balseiro, D. Feinberg, and M. Avignon, Phys. Rev. Lett. \textbf{101} 107001 (2008).

\bibitem{Mironov_SOC} S. V. Mironov, A. S. Mel’nikov, A. I. Buzdin, Phys. Rev. Lett. \textbf{114}, 227001 (2015).

\bibitem{Bergeret_SOC} F. Konschelle, I. V. Tokatly,  F. S. Bergeret, Phys. Rev. B \textbf{92}, 125443 (2015).

\bibitem{Martin} A. Zazunov, R. Egger, T. Jonckheere, T. Martin, Phys. Rev. Lett. \textbf{103}, 147004 (2009).

\bibitem{Martin_2} L. Dell’Anna, A. Zazunov, R. Egger, T. Martin, Phys. Rev. B \textbf{75}, 085305 (2007).

\bibitem{Brunetti} A. Brunetti, A. Zazunov, A. Kundu, R. Egger, Phys. Rev. B \textbf{88}, 144515 (2013).

\bibitem{Nazarov_PRB} T. Yokoyama, M. Eto, and Yu. V. Nazarov, Phys. Rev. B \textbf{89}, 195407 (2014).

\bibitem{Campagnano} G. Campagnano, P. Lucignano, D. Giuliano, A. Tagliacozzo, J. Phys. Condens. Matter \textbf{27}, 205301 (2015).

\bibitem{Kouwenhoven} D. B. Szombati, S. Nadj-Perge, D. Car, S. R. Plissard, E. P. A. M. Bakkers, and L. P. Kouwenhoven, Nature Phys. \textbf{12}, 568 (2016).

\bibitem{Nesterov} K. N. Nesterov, M. Houzet, J. S. Meyer, Phys. Rev. B \textbf{93}, 174502 (2016).

\bibitem{Ying} Z.-J. Ying, M. Cuoco, P. Gentile, and C. Ortix, \textit{2017 16th International Superconductive Electronics Conference (ISEC)}, Naples, Italy (2017). 

\bibitem{Spanslatt} C. Sp\r{a}nsl\"{a}tt, Phys. Rev. B \textbf{98}, 054508 (2018).

\bibitem{Kutlin} A. G. Kutlin, A. S. Mel'nikov
Phys. Rev. B \textbf{101}, 045418 (2020).

\bibitem{Kopasov} A. A. Kopasov, A. G. Kutlin, and A. S. Mel'nikov, Phys. Rev. B \textbf{103}, 144520 (2021).


\bibitem{GurlichPRB2010} C. G\"{u}rlich, S. Scharinger, M. Weides, H. Kohlstedt, R. G. Mints, E. Goldobin, D. Koelle, and R. Kleiner, Phys. Rev. B \textbf{81}, 094502 (2010).

\bibitem{SickingerPRL2012} H. Sickinger, A. Lipman, M. Weides, R. G. Mints, H. Kohlstedt, D. Koelle, R. Kleiner, and E. Goldobin, Phys. Rev. Lett. \textbf{109}, 107002 (2012).

\bibitem{UstinovAPL2002} A. V. Ustinov, Appl. Phys. Lett. \textbf{80}, 3153 (2002).
\bibitem{GaberPRB2005} T. Gaber, E. Goldobin, A. Sterck, R. Kleiner, D. Koelle, M. Siegel, and M. Neuhaus, Phys. Rev. B \textbf{72}, 054522 (2005).
\bibitem{GoldobinPRB2016} E. Goldobin, S. Mironov, A. Buzdin, R. G. Mints, D. Koelle, and R. Kleiner, Phys. Rev. B \textbf{93}, 134514 (2016).


\bibitem{Pickett} W. E. Pickett and J. S. Moodera, Phys. Today \textbf{54}(5), 39 (2001).

\bibitem{Coey} J. M. D. Coey and M. Venkatesan, J. Appl. Phys. \textbf{91}, 8345 (2002).

\bibitem{Keizer} R. S. Keizer, T. B. Goennenwein, T. M. Klapwijk, G. Miao, G.
Xiao, and A. Gupta, Nature (London) \textbf{439}, 825 (2006).

\bibitem{Anwar} M. S. Anwar, F. Czeschka, M. Hesselberth, M. Porcu, and J. Aarts, Phys. Rev. B \textbf{82}, 100501(R) (2010).



\bibitem{Eschrig_1} M. Eschrig, T. L\"{o}fwander, T. Champel, J. C. Cuevas, J. Kopu,
and G. Sch\"{o}n, J. Low Temp. Phys. \textbf{147}, 457 (2007).

\bibitem{Eschrig_2} M. Eschrig and T. L\"{o}fwander, Nat. Phys. \textbf{4}, 138 (2008).

\bibitem{Eschrig_3} R. Grein, M. Eschrig, G. Metalidis, and G. Sch\"{o}n, Phys. Rev. Lett. \textbf{102}, 227005 (2009).

\bibitem{Eschrig_4} M. Eschrig, A. Cottet, W. Belzig, and J. Linder, New J. Phys. \textbf{17}, 083037 (2015).

\bibitem{Mironov_HM} S. Mironov, A. Buzdin, Phys. Rev. B \textbf{92}, 184506 (2015).

\bibitem{Meng_Flux} S. Mironov, H. Meng, A. Buzdin, Appl. Phys. Lett. \textbf{116}, 162601 (2020). 
	
\bibitem{Devizorova} Zh. Devizorova and S. Mironov, Phys. Rev. B \textbf{95}, 144514 (2017).


\bibitem{Margaris} I.~Margaris, V.~Paltoglou, and N~Flyzanis, J. Phys.: Condens. Matter \textbf{22}, 445701 (2010).

\bibitem{Halterman} K.~Halterman, M.~Alidoust, R.~Smith, and S.~Starr, Phys. Rev. B \textbf{105}, 104508 (2022).

\bibitem{KalenkovPRB2009} M.~S.~Kalenkov, A.~V.~Galaktionov, and A.~D.~Zaikin, Phys. Rev. B \textbf{79}, 014521 (2009).

\bibitem{BeenakkerPRL1991} C.~W.~J.~Beenakker, Phys. Rev. Lett. \textbf{67}, 3836 (1991).

\bibitem{Devizorova_2} Zh. Devizorova, S. Mironov, A. I. Buzdin, Phys. Rev. B \textbf{103}, 224510 (2021).

\bibitem{PGdeGennes} P. G. de Gennes, Superconductivity of Metals and
Alloys, Benjamin, New York, 1966 (Chap.5).

\bibitem{Buz} A. I. Buzdin, Rev. Mod. Phys. \textbf{77}, 935 (2005).

\bibitem{Shukrinov} Y. M. Shukrinov, I. R. Rahmonov, K. Sengupta, A. Buzdin, Appl. Phys. Lett. \textbf{110}, 182407 (2017).
 
\bibitem{Visani} C. Visani, Z. Sefrioui, J. Tornos, C. Leon, J. Briatico, M. Bibes, A.
Barthlmy, J. Santamara, J. E. Villegas, Nat. Phys. \textbf{8}, 539 (2012).
 
\bibitem{Sanchez-Manzano} D. Sanchez-Manzano, S. Mesoraca, F. A. Cuellar, M. Cabero, V. Rouco, G. Orfila, X. Palermo, A. Balan, L. Marcano, A. Sander, M. Rocci, J. Garcia-Barriocanal, F. Gallego, J. Tornos, A. Rivera, F. Mompean, M. Garcia-Hernandez, J. M. Gonzalez-Calbet, C. Leon, S. Valencia, C. Feuillet-Palma, N. Bergeal, A. I. Buzdin, J. Lesueur, Javier E. Villegas, J. Santamaria, Nat. Mater. \textbf{21}, 188 (2022).

\bibitem{Jungxiang} Y. Jungxiang, R. Fermin, K. Lahabi, J. Aarts, arXiv:2303.13922 (2023).

\bibitem{Birge_1} J. A. Glick, V. Aguilar, A. B. Gougam, B. M. Niedzielski, E. C. Gingrich, R. Loloee, W. P. Pratt Jr., N. O. Birge, Sci. Adv. \textbf{4}, eaat9457 (2018).

\bibitem{Birge_2} V. Aguilar, D. Korucu, J. A. Glick, R. Loloee, W. P. Pratt, Jr., N. O. Birge, Phys. Rev. B \textbf{102}, 024518 (2020).


\end{thebibliography}
\end{document}